%% file: main.tex
\documentclass[conference,final]{IEEEtran}
\IEEEoverridecommandlockouts
\usepackage{cite}
\usepackage{amsmath,amssymb,amsfonts}
\usepackage{graphicx}

\usepackage{algorithm,algpseudocode}
\usepackage{textcomp}
\usepackage{xcolor}
\usepackage{hyperref}
\usepackage{amsmath}
\usepackage[numbers]{natbib}
\usepackage{booktabs}
\usepackage{bm}
\usepackage{subfigure} 
\newtheorem{theorem}{Theorem}[section]
\newtheorem{lemma}{Lemma}[section]

\def\BibTeX{{\rm B\kern-.05em{\sc i\kern-.025em b}\kern-.08em
    T\kern-.1667em\lower.7ex\hbox{E}\kern-.125emX}}
\begin{document}

\input{math_commands.tex}

\newcommand{\our}{{Sim2Rec}}
\newcommand{\didi}{{DidiChuxing}}

\title{\our{}: A Simulator-based Decision-making Approach to Optimize Real-World Long-term User Engagement in Sequential Recommender Systems}

\author{
\IEEEauthorblockN{Xiong-Hui Chen$^{1,3}$, Bowei He$^{5}$, Yang Yu$^{1,3, 4, *}$\thanks{ $^*$ Corresponding author},Qingyang Li$^{2}$, Zhiwei Qin$^{6,+}$\thanks{ $^+$ Work done while the author was with DiDi AI Labs}, Wenjie Shang$^{2}$, Jieping Ye$^{2}$, Chen Ma$^{5}$}
\IEEEauthorblockA{\textit{$^{1}$ National Key Laboratory of Novel Software Technology, Nanjing University, Nanjing, China}\\
  \textit{$^{2}$ DiDi AI Labs,
  $^{3}$ Polixir.ai, $^{4}$ Peng Cheng Laboratory, Shenzhen, China,  
  $^{5}$ City University of Hong Kong, $^{6}$ Lyft}\\
chenxh@lamda.nju.edu.cn, boweihe2-c@my.cityu.edu.hk, yuy@nju.edu.cn, qingyangli@didiglobal.com, zq2107@caa.columbia.edu, \\
shangwenjie@didiglobal.com, jieping@gmail.com, chenma@cityu.edu.hk
\vspace{-4mm}
}
}
\maketitle
\begin{abstract}    
	Long-term user engagement  (LTE) optimization in sequential recommender systems (SRS) is shown to be suited by reinforcement learning (RL) which finds a policy to maximize long-term rewards.  Meanwhile, RL has its shortcomings, particularly requiring a large number of online samples for exploration, which is risky in real-world applications. 
	One of the appealing ways to avoid the risk is to build a simulator and learn the optimal recommendation policy in the simulator.  In LTE optimization, the simulator is to simulate multiple users' daily feedback for given recommendations. 
	However, building a user simulator with no reality-gap, i.e., can predict user's feedback exactly, is unrealistic because the users' reaction patterns are complex and historical logs for each user are limited, which might mislead the simulator-based recommendation policy.
	In this paper, we present a practical simulator-based recommender policy training approach, Simulation-to-Recommendation (\our{}) to handle the reality-gap problem for LTE optimization.  Specifically, \our{} introduces a simulator set to generate various possibilities of user behavior patterns, then trains an environment-parameter extractor to recognize users' behavior patterns in the simulators.  Finally, a context-aware policy is trained to make the optimal decisions on all of the variants of the users based on the inferred environment-parameters.  
	The policy is transferable to unseen environments (e.g., the real world)  directly as it has learned to recognize all various user behavior patterns and to make the correct decisions based on the inferred environment-parameters.  
    Experiments are conducted in synthetic environments and a real-world large-scale ride-hailing platform, \didi{}. The results show that \our{} achieves significant performance improvement, and produces robust recommendations in unseen environments.
\end{abstract}

\begin{IEEEkeywords}
reinforcement learning, reality gaps, recommender systems
\end{IEEEkeywords}

\input{introduction.tex}

\input{relatedwork.tex}

\input{preli.tex}

\input{method.tex}

\input{experiment.tex}

	  \vspace{-2mm}
	\section{Discussion and Future Work} \label{sec:conc}

    
    In this work, we study the reality-gap problem of simulator-based RL for LTE optimization in SRS. We first formulate the problem based on the zero-shot policy transfer framework and propose the extra challenges of solving the reality-gap problem on SRS. We build a practical Simulation To Recommendation (\our{}) algorithm to handle the above challenges to give a reliable policy in the real world. 
    The experiments are conducted in a synthetic environment and a real-world application. We use a synthetic environment to quantify the performance improvement of the proposed environment-parameter extractor. In the real-world application, we verify the necessity of the proposed techniques and the effectiveness of the proposed method in the production environment. 
    
    Simulator-based RL is a promising way to avoid trial-and-error costs to learn policies in real-world sequential recommender systems. We hope the reasonable performance of \our{} will inspire researchers to develop more powerful recommender systems by handling the reality-gaps. 
    The limitation of current \our{} mainly comes from the implementation of the proposed techniques in Sec.~\ref{sec:reliable}, which are designed only based on empirical techniques. We believe that more theoretical solutions to solve the problems, e.g., uncertainty evaluation and extrapolation error evaluation, can be further discussed, which will be in our future work.

\vspace{-1mm}
\section*{Acknowledgements}

This work is supported by the National Key Research and Development Program of China (2020AAA0107200), the National Science Foundation of China (61921006) and the Major Key Project of PCL (PCL2021A12).

\bibliographystyle{IEEEtranN}
\bibliography{main-ref}
\clearpage
\onecolumn
\appendix

\subsection{Proof}
\label{app:proof}

We give the proof of Lemma~\ref{lemma:elbo} and Theorem~\ref{the:1-main}:
\paragraph{Proof of Lemma~\ref{lemma:elbo}} The objective of Eq.~\eqref{equ:kld-main} can be rewritten as:
	\begin{align}
		& KLD(q_\kappa(\upsilon\mid X)\Vert p_\theta(\upsilon\mid X)) \label{equ:obj2} \\
		=&\mathbb{E}_{q_\kappa(\upsilon\mid X)} \left[ \log \frac{q_\kappa(\upsilon\mid X)}{p_\theta(\upsilon\mid X)} \right]\notag\\
		=&\mathbb{E}_{q_\kappa(\upsilon\mid X)} \left[ \log q_\kappa(\upsilon\mid X) - \log  p_\theta(\upsilon, X) + \log p_\theta(X) \right]\notag\\\
		=&-L(\theta, \kappa; X)  + \log p_\theta(X). \label{equ:obj}
	\end{align}
	Since $\log p_\theta(X)$ is independent of $q_\kappa(\upsilon\mid X)$, minimizing Eq.~\eqref{equ:obj2} is equivalent to maximizing $L(\theta, \kappa; X)$ in Eq.~\eqref{equ:obj}. 	Based on Bayes's theorem, we have: 
	\begin{small}
	\begin{align}
		L(\theta, \kappa; X) =&\mathbb{E}_{q_\kappa(\upsilon\mid X)} \left[ -\log q_\kappa(\upsilon\mid X) +\log  p_\theta(\upsilon, X)  \right] \notag\\
		=&\mathbb{E}_{q_\kappa(\upsilon\mid X)} \left[ -\log q_\kappa(\upsilon\mid X) + \log \left( p_\theta(X\mid \upsilon)p_\theta(\upsilon) \right) \right] \notag \\
		=&\mathbb{E}_{q_\kappa(\upsilon\mid X)} \left[ \log \frac{p_\theta(\upsilon)}{q_\kappa(\upsilon\mid X)} +  \log p_\theta(X\mid \upsilon) \right]\notag \\
		=&\mathbb{E}_{q_\kappa(\upsilon\mid X)} \left[\log p_\theta(X\mid \upsilon) \right] - KLD \left( q_\kappa(\upsilon\mid X)\Vert p_\theta(\upsilon) \right). \notag
	\end{align}
	\end{small}
	Under the assumption that $X$ is i.i.d. sampled from $\dataset$, we obtain the evidence lower bound (ELBO) objective:
	\begin{small}
	$$\max_{\kappa,\theta} \mathbb{E}_{X\sim\dataset}\left[ \mathbb{E}_{q_\kappa(\upsilon\mid X)}\left[\log p_\theta(X\mid \upsilon) \right] - KLD\left(q_\kappa(\upsilon\mid X) \Vert p(\upsilon) \right) \right]. $$
	\end{small}
\paragraph{Proof of Theorem~\ref{the:1-main}}
	In the RL scenario, the action is sampled conditionally on the state, thus the posterior $p_\theta$ can be separated by: 
	\begin{align}
		&p_\theta\left(s^{(i)}, a^{(i)}\mid \upsilon\right) \notag\\ 
		= &p_\theta\left(a^{(i)}\mid \upsilon, s^{(i)}\right) p_\theta\left(s^{(i)}\mid \upsilon\right) \notag\\
		= &p_{\psi_a}\left(a^{(i)}\right) p_\theta\left(\psi_a\mid \upsilon, s^{(i)}\right) p_{\psi_s}\left(s^{(i)}\right)  p_\theta\left(\psi_s\mid \upsilon\right)
		\label{equ:recons-rl},
	\end{align}
	where $\psi_s$ and $\psi_a$ denote the decoded parameters of the distribution. Based on Lemma~\ref{lemma:elbo}, the tractable objective of Eq.~\ref{equ:kld-main} can be written as:
	\begin{small}
		\begin{align}
			&  \mathbb{E}_{X\sim\dataset, q_\kappa(\upsilon\mid X)}\left[\log p_\theta(X\mid \upsilon) \right] - KLD\left(q_\kappa(\upsilon\mid X) \Vert p(\upsilon) \right) \notag \\
			= &  \mathbb{E}_{X\sim\dataset,q_\kappa(\upsilon\mid X)}\left[ \sum_{i}^{N} \log p_\theta\left(x^{(i)}\mid \upsilon\right)  \right] - KLD\left(q_\kappa(\upsilon\mid X) \Vert p(\upsilon) \right) \notag\\
			\begin{split}
				= &  \mathbb{E}_{X\sim\dataset,q_\kappa(\upsilon\mid X)}\left[\sum_{i=1}^N  \log p_\theta\left(s^{(i)}\mid \upsilon\right) + \log p_\theta\left(a^{(i)}\mid \upsilon, s^{(i)}\right) \right] \\
				& - KLD\left(  q_\kappa(\upsilon\mid X) \Vert p(\upsilon) \right). \notag
			\end{split}
		\end{align}	
	\end{small}
	$q_\kappa\left(\upsilon\mid s^{(i)}, a^{(i)}\right)$ can be modeled with Gaussian distribution, then the result  $q_\kappa\left(\upsilon\mid X \right)$ is also a Gaussian distribution with a closed-form solution~\cite{RakellyZFLQ19}. For any differentiable $p_{\psi_s}$ and $p_{\psi_a}$, the ELBO objective is tractable via the reparameterization trick~\citep{KingmaW13}.

\clearpage
\subsection{Visualization of PCA}
\label{app:pca}
\begin{figure*}[h]
	\centering
	\hspace{-10mm}
	\subfigure[epoch 0]{
		\includegraphics[width=0.19\linewidth]{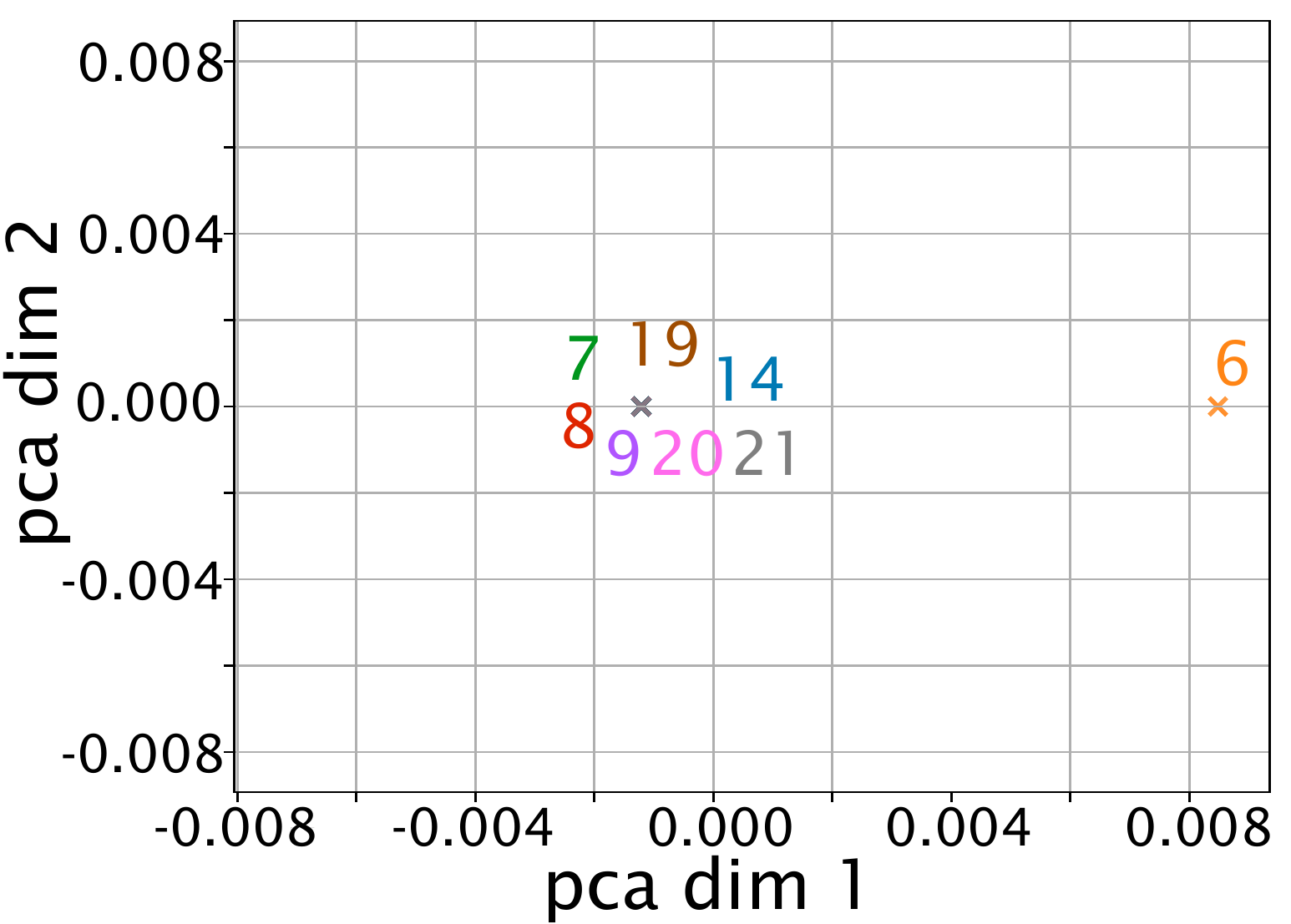}
	}
	\subfigure[epoch 2000]{
	\includegraphics[width=0.19\linewidth]{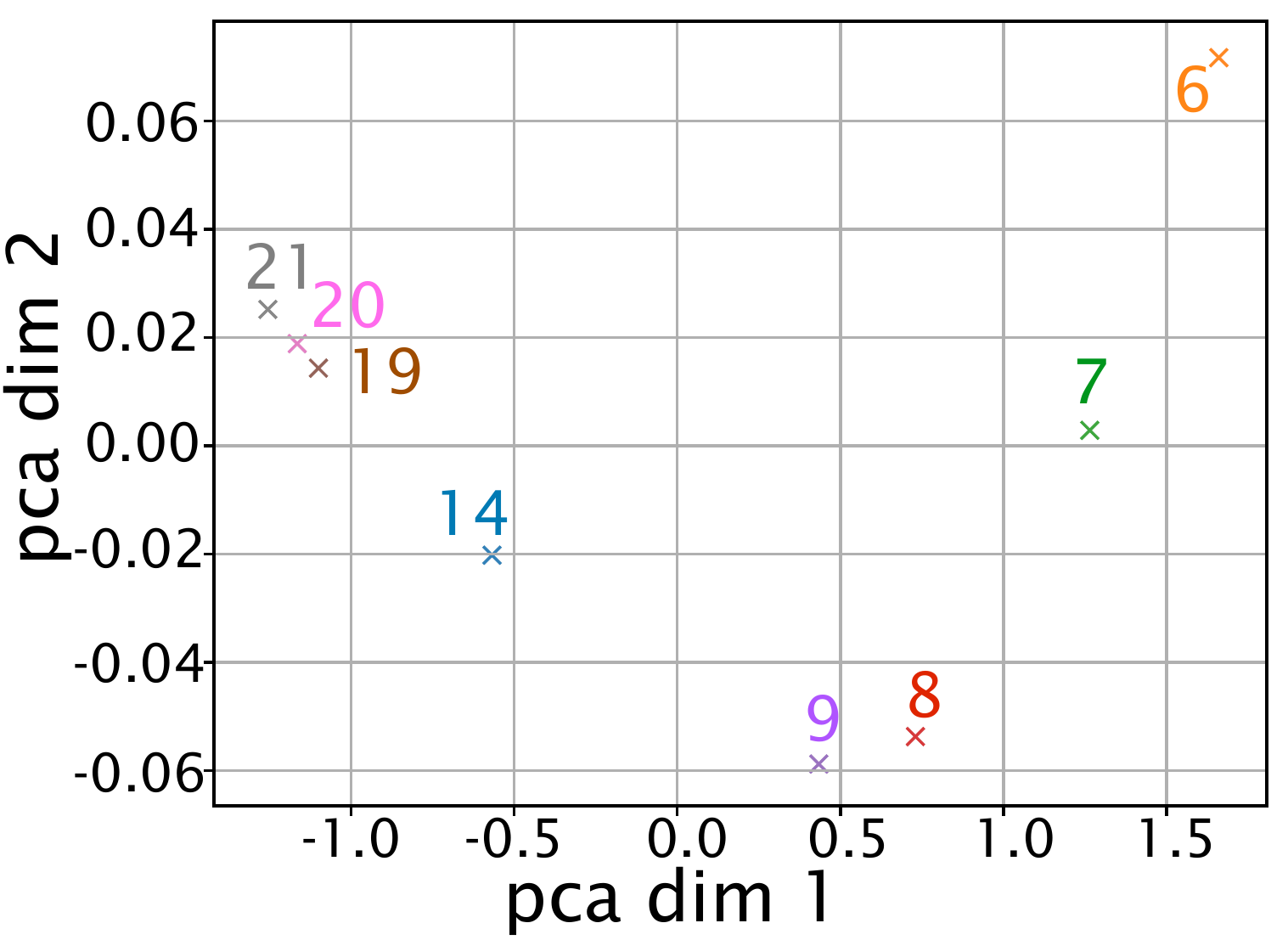}
	}	
	\subfigure[epoch 4000]{
	\includegraphics[width=0.19\linewidth]{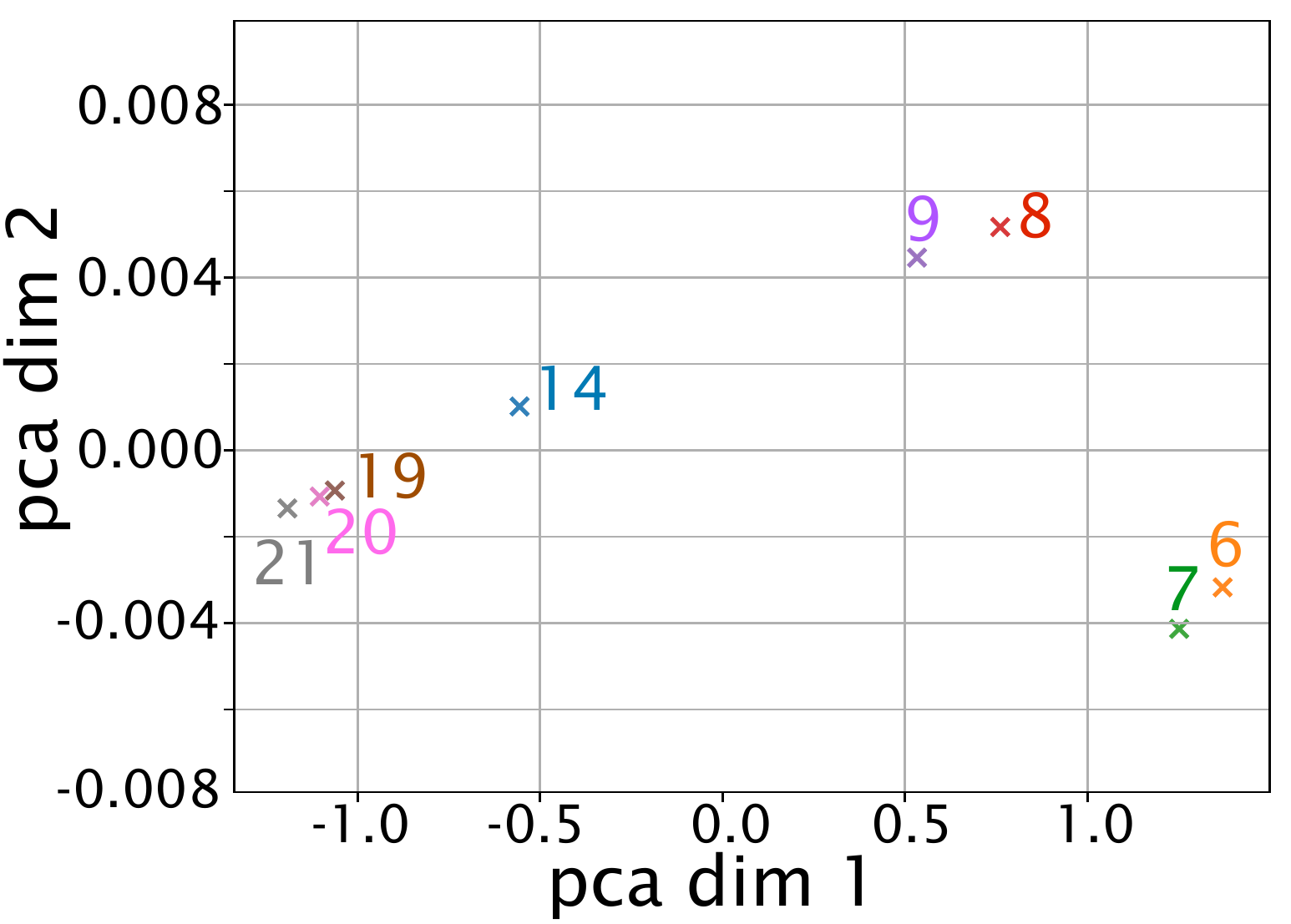}
	}
	\subfigure[epoch 6000]{
	\includegraphics[width=0.19\linewidth]{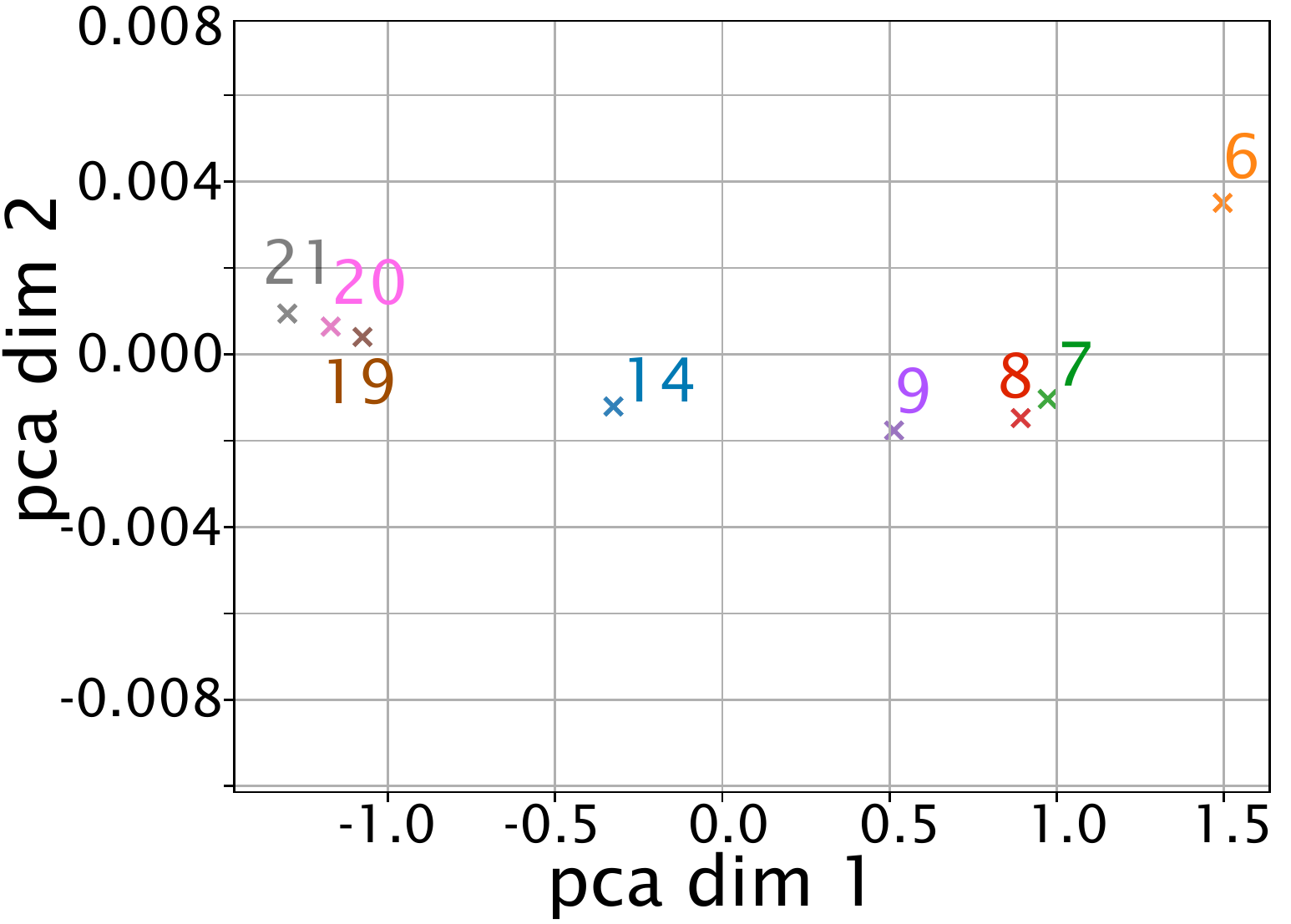}
	}
	\subfigure[epoch 8000]{
	\includegraphics[width=0.19\linewidth]{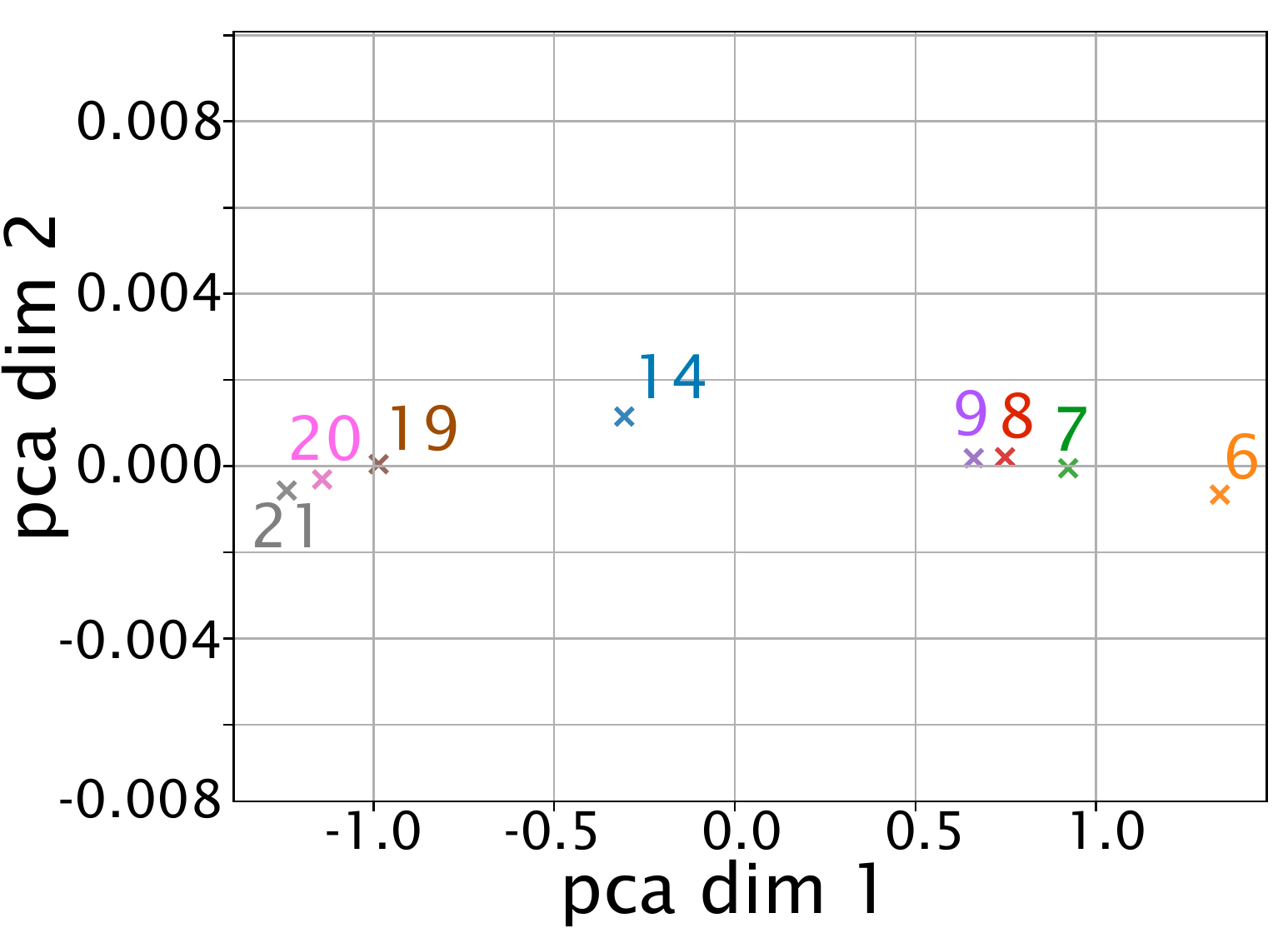}
	}
	\caption{Illustration of the visualization on $\upsilon$. The X-axis denotes the first principal component, and the Y-axis denotes the second one. Each cross point denotes the projection of the latent code for the state distribution. The numbers with the same color to the point denote the ground-truth environment parameter $\omega_\group$. Since $q_\kappa(\upsilon\mid X)$ is a Gaussian distribution, we only draw the mean of the distribution for legibility.}
	\label{fig:lts:pca-vis}
\end{figure*}

\end{document}

%% file: math_commands.tex
\newcommand{\dataset}{\mathcal{D}}
\newcommand{\sspac}{\mathcal{S}}
\newcommand{\aspac}{\mathcal{A}}
\newcommand{\discount}{\gamma}
\newcommand{\occ}{\rho}
\newcommand{\taj}{\tau}
\newcommand{\T}{T}
\newcommand{\acs}{a}
\newcommand{\state}{s}
\newcommand{\rew}{R}
\newcommand{\trans}{P}
\newcommand{\ts}{t}
\newcommand{\initdist}{d_0}

\newcommand{\gspac}{\mathcal{G}}
\newcommand{\uspac}{\mathcal{U}}
\newcommand{\ispac}{\mathcal{I}}
\newcommand{\fbspac}{\mathcal{Y}}
\newcommand{\fb}{y}
\newcommand{\user}{u}
\newcommand{\group}{g}
\newcommand{\itm}{i}

\newcommand{\suser}{s^{\rm user}}
\newcommand{\shist}{s^{\rm hist}}
\newcommand{\sstat}{s^{\rm stat}}
\newcommand{\sgroup}{s^{\rm group}}
\newcommand{\stime}{s^{\rm time}}

\newcommand{\sruser}{s^{\rm user,r}}
\newcommand{\srhist}{s^{\rm hist,r}}
\newcommand{\srstat}{s^{\rm stat,r}}
\newcommand{\srgroup}{s^{\rm group,r}}
\newcommand{\srtime}{s^{\rm time,r}}
\newcommand{\rtraj}{\tau^{\rm r}}
\newcommand{\realworld}{\mathcal{E}}
\newcommand{\alg}{\mathcal{H}}

%% file: introduction.tex
\vspace{-3mm}
\section{Introduction}

Sequential Recommender Systems (SRS) that aim to recommend potentially relevant item sequences for users have played an important role in various internet platforms like ride-hailing apps~\cite{demer,ride1}, E-commerce sites~\cite{virtualtaobao,ecom1,ecom2,ecom3}, and videos sites~\cite{video1,video2}. Increasing the long-term engagement (LTE),  typically representing users' desire to stay and keep active in the platforms, is an critical objective in SRS~\cite{video1,rl-rs-lte}. Recent studies have shown that reinforcement learning (RL) is a promising approach for optimizing LTE.  They treat the recommendation procedures as sequential interactions between users and a recommender agent~\cite{usersim}, then use RL to find an optimal policy that maximizes cumulative rewards of users from the interactions.

However, RL methods rely on a large number of trial-and-error samples in the real world, which obstruct the further applications of RL in those risk-sensitive platforms~\cite{offline-overview,chen2019reinforcement}. 
Training RL policy in a simulator is an ideal way to avoid trial-and-error costs. In SRS scenarios,  a simulator is to simulate users’ responses to given recommendations. 
However, building an accurate simulator is unrealistic, since user behaviors are often complex and the historical logs are limited~\cite{rl-rs-lte}.  
The discrepancy between simulation and reality, referred to as the reality-gaps, results in undesired real-world  performance degradation of the policies learned from standard RL paradigm~\cite{realitygap}.   However, in SRS, the ill-posedness of the standard RL paradigm based on the simulator with reality-gaps has rarely been discussed explicitly.

In this paper, we focus on handling the reality-gaps problem of simulator-based RL for LTE optimization.  We introduce zero-shot policy transfer techniques based on an environment-parameter extractor for SRS to handle the problem.
Zero-shot policy transfer techniques have been widely used to overcome the reality-gaps of physical simulators in challenging tasks ~\cite{osi-lstm,OpenAIcube,dr}. These techniques assume the reality-gaps come from the gap of environment parameters (e.g., different friction coefficients for robot control). They first construct a simulator set with a massive number of different environment parameters selected from the environment-parameter space. Based on the simulator set, they learn an extractor to infer the environment parameters from interaction trajectories, and a context-aware policy to control an agent to perform adaptable behaviors for optimal performance according to the inferred parameters \cite{osi-lstm,OpenAIcube}.  When deployed, the extractor adjusts its inferred environment parameters via the real interaction trajectory information and thus adapts the policy to suitable behaviors automatically. 
Policy transfer is completed after the policy collects enough samples and the extractor determines the environment parameters. If the environment-parameter space covers the environment parameters of the real world and the simulator set has traversed the space, we can claim that the extractor can infer the correct parameters and the policy will be adapted to make correct decisions.

However, SRS scenarios are different from the existing applications of zero-shot policy transfer in the following aspects: 
First, in SRS scenarios, a policy serves multiple users in multiple regions at the same time. 
The environment-parameter extractor should identify the behavior pattern of each user. Besides, each region also has its own context, leading to inconsistency in user behaviors among different regions. For instance, in ride-hailing platforms, drivers in different cities may have different engagements (e.g., online time), independent of their personas, since the base number of passengers is not in the same order of magnitude in these cities. The behavioral differences among regions are referred to as group-behavior differences in this study, which is common in the real world~\citep{group-behvaior}. 
In this scenario, the representation of environment parameters would be hard to identify if merely considering a single user's interaction trajectory. 
Second, in SRS scenarios, the user simulator is hard to model by ``physical rules'', thus current practical algorithms learn to simulate from data~\citep{demer,simulator-mayi} through neural networks. In this scenario, the environment-parameter space is the weight space of neural networks, which is extremely large and redundant. It is almost impractical to develop an extractor and a policy to identify the environment parameters in such a space.

In this work, we first formulate the reality-gaps based on the concept of environment parameters and analyze the extra challenges of the reality-gaps. Based on the analysis, we build a new zero-shot policy transfer system, named \textbf{Sim}ulation-to-\textbf{Rec}ommendation (\our{}), which handles the reality-gaps through an environment-parameter extractor.  To solve the environment-parameters extraction problem in SRS, we propose a hierarchical environment-parameter extractor, which includes an \textbf{S}tate-\textbf{A}ction \textbf{D}istributional
variational \textbf{A}uto\textbf{E}ncoder (SADAE), based on the theoretical analysis of evidence lower bound, to embed a state-action dataset of a user group into a latent vector. Based on the embedded group-information vector, we use a recurrent neural network (RNN)~\citep{gru} to identify the parameters of each user; 
To handle the problem of extremely large and redundant environment-parameter space of the data-driven user simulator, we develop several techniques for using the simulator and policy exploration to keep the feasibility of the framework in real-world SRS applications.

	In summary, the main contributions of this paper are:
	\begin{itemize}
	    \item To handle the reality-gap problem of simulator-based RL methods in SRS, we propose a zero-shot policy transfer approach, Sim2Rec. To the best of our knowledge, this is the first work that considers the reality-gaps of the simulator in policy optimization for SRS;
	    \item To identify the environment-parameter efficiently in the SRS scenario, we propose a hierarchical environment-parameter extractor architecture which includes a new autoencoder SADAE to embed a state-action dataset of a user group into a latent vector. Several techniques are introduced to reduce the environment-parameter space of the data-driven simulator into a feasible scale to facilitate the policy and extractor learning;
		\item We conduct experiments in an open-source synthetic environment and a real-world ride-hailing platform, \didi{}. The results in synthetic environments, offline tests, and online deployment demonstrate the effectiveness of \our{}.
	\end{itemize}

%% file: relatedwork.tex
\section{Related Work}

	Training RL policy in a simulator is an ideal way to avoid costly trial-and-errors in the real environment~\citep{luo2022survey}.  Many RL-based SRS approaches regarded the simulator as the oracle environment for training and testing~\citep{matrix-factorization-model,RecSim}. 
	Recent studies focus on data-driven simulator reconstruction with different methods: \cite{demer,galileo} use a generative adversarial framework to learn a simulator to generate a data distribution consistent with the real distribution; \cite{pseudo-dyna-q} construct a simulator via a World Model;  \citet{zhu2022offline} improve the generalization ability of the world model through causal Structured model.
	\cite{debiase-1} use inverse propensity weighting techniques to handle the selection bias problem to construct a debiased simulator.
    \citet{s2r_rec} use a real dataset to correct the representation and reward function of a simulator to improve the fidelity.
	In real applications, it is inevitable that reconstructed simulators have reality-gaps since customer behaviors are often highly complex. 
	However, current studies does not considers the reality-gaps of the simulators when learning a RL policy, which might results in undesired real-world performance~\cite{realitygap}.

	On the other hand, zero-shot policy transfer techniques have been widely used to overcome the reality-gaps of physical simulators in challenging tasks~\cite{osi-lstm,OpenAIcube,dr,SadeghiL17,dr1,lee_2020,pearl,luo2022adapt}.  These techniques use physical simulators, which are built by human experts based on laws of physics, for policy learning and assume the reality-gaps come from the errors of environment parameters estimation (e.g., friction coefficients for robot control) of the simulators.  The paradigm of zero-shot policy transfer techniques can be summarized into two phases: (1) construct a simulator set with a massive number of different environment parameters selected from the environment-parameter space; (2) train a policy that can take reasonable actions in the simulator set. If the environment-parameter space covers the environment parameters of the real world and the simulator set has traversed the space, we can claim that, when deployed, the policy can make reasonable decisions in the real world as in the simulators. 
	One popular way to learn the policy is learning/constructing an online system identification (OSI) module~\cite{osi-lstm,OpenAIcube,zhou_2019,pearl,learn_to_adapt}  to infer the environment parameters from interaction trajectories, and a context-aware policy to control an agent to perform adaptable behaviors for optimal performance according to the inferred parameters.  
	When deployed, the OSI module adjusts its inferred environment parameters via the real interaction trajectories and thus adapts the policy to suitable behaviors automatically. \cite{zhou_2019} design an EPI-policy to probe some interaction trajectories, an EPI-trajectory-embedding network for environment-parameter representation which can predict the dynamics of the corresponding simulator, and a task-specific policy to perform optimal behaviors based on the inferred representations of each simulator. 
	\cite{osi-lstm,OpenAIcube,learn_to_adapt,luo2022adapt} use a end-to-end architecture for environment-parameter representation and adaptable policy learning. A recurrent neural network (RNN), e.g., LSTM~\cite{lstm}, is introduced for environment-parameter representation, then the context-aware policy takes actions based on the outputs of RNN and the current states. 
	In this work, we follow the basic idea of zero-shot policy transfer and the end-to-end architecture as previous. We formulate and analyze the extra challenges of the standard zero-shot policy transfer framework for SRS, and proposed a practical solution to handle these challenges.

%% file: preli.tex
\section{Problem Formulation}

We first formulate the general workflow of SRS.  In SRS, a recommendation system serves multiple users $\user \in \uspac$ in multiple groups $\group \in \gspac$, $\uspac$ and $\gspac$ are the user and group space respectively. A recommendation policy $\pi$ interacts with those users at discrete time steps $\ts \in \{0, 1, ..., \T\}$ within a recommendation session, where $\T$ is the maximal time steps of a recommendation session. At each time-step $\ts$, the policy $\pi$ will give each user an item $\itm \in \ispac$ and receive a feedback $\fb \in \fbspac$ from each user, where $\ispac$ is the item space and $\fbspac$ is the feedback space.
Taking the ride-hailing platform as an example, the platform provides services in multiple cities (i.e., groups $\group$), and interacts with numerous drivers  (i.e., users $\user$) in each city. The system will design several program items $\itm$ to recommend. Each program item includes a task for the driver to follow, e.g., a dispatch task that guides the driver to some regions.  The platform will receive the driver's feedback $\fb$ like fulfilling some orders or just going offline.

\subsection{Markov Decision Process Formulation}
RL-based recommender systems treat the recommendation task as sequential interactions between a recommender system (agent) and users (environment),and use a Markov Decision Process (MDP) to model them~\cite{usersim,sutton2018rl,rl-rs-lte}.  A MDP is defined by a tuple of five elements $(\sspac, \aspac, \rew, \trans, \discount,\initdist)$, where $\sspac$ and $\aspac$ is the state space and action space respectively, $\trans: \sspac \times \aspac \rightarrow \sspac$ is the transition function, $\rew: \sspac \times \aspac \times \sspac \rightarrow \mathbb{R}$ is the mean reward function, $\discount \in [0,1]$ is the discount factor and $\initdist$ is the initial state distribution. A recommendation policy $\pi: \sspac \rightarrow \aspac$. For LTE optimization, $(\sspac, \aspac, \rew, \trans)$ are set as follow:
\begin{itemize}
	\item \textbf{State space $\sspac$}: The state is composed of these parts: user profile feature $\suser$ (e.g., age, gender, and location), user’s history of feedback  $\shist$ (e.g., number of order fulfilling and online time) and their statistics  $\sstat$ (e.g., averaged number of order fulfilling in recent 7 and 14 days), some external features of the group $\sgroup$ where the user in (e.g., city information), and some timestep related features  $s^{\rm time}$ (e.g., weather). 
	\item \textbf{Action space $\aspac$}: Instead of letting $a$ as the index of the items in $\ispac$~\cite{rl-rs-lte,video1}, we formulate action $\acs$ as a set of parameters that can determine the recommended item $\itm$ from $\ispac$, which is the same as previous studies like \cite{demer,virtualtaobao}. Specifically, we have a predefined rule-based function $F: \aspac \rightarrow \ispac$. For example, for each timestep $\ts$, $\pi(\acs_\ts|\state_\ts)$ determines the difficulty coefficient of tasks $\acs_\ts$ for each driver, then $F(\acs_\ts)$ finds the corresponding program item $\itm_\ts$ to recommend to the driver.
	\item \textbf{Reward function $\rew$}: For each time-step $\ts$, we define a metric of instant engagement $\rew(\state_\ts,\acs_\ts,\state_{\ts+1})$ through the current state $\state_\ts$, taken action $\acs_\ts$ and user feedback (in $s_{\ts+1}$). Then we define the metric of LTE as $\sum_{\ts=0}^\T \rew(\state_\ts,\acs_\ts,\state_{\ts+1})$ and ignore the delayed metrics~\cite{rl-rs-lte} for problem simplification. 
	\item \textbf{Transition function $\trans$}: $\trans(\state_{\ts+1} | \state_\ts,\acs_\ts)$ defines the state transition from $\state_\ts$ to $\state_{\ts+1}$ after taking action $\acs_\ts$.
\end{itemize}

\subsection{Simulator-based RL for LTE Optimization}

In this article, we follow a general pipeline of simulator-based RL for LTE optimization as~\cite{usersim,demer,virtualtaobao}.  We first define a user simulator $M: \sspac \times \aspac \times \sspac \rightarrow \fbspac$. Specifically, the goal of a user simulator can be formally defined as follows: given a state-action pair $(\state, \acs)$, imitate  the  user’s feedback (behavior) $\fb$ on a recommended action $\acs$ according to the state $\state$.  For each timestep $\ts$, given predicted $\hat y_{\ts+1}$, we first update $\shist_{\ts+1}$ and $\sstat_{\ts+1}$ through $\hat y_{\ts+1}$, then load $\sruser_{\ts+1}$, $\srgroup_{\ts+1}$, and $\srtime_{\ts+1}$ from a real trajectory $\tau^r$ in logged dataset $\dataset$, where $\rtraj:=[s^r_0, a^r_0, s^r_1, a^r_1, ..., s^r_T, a^r_T]$. Finally, we have $\state_{\ts+1} = [\shist_{\ts+1},\sstat_{\ts+1},\sruser_{\ts+1},\srgroup_{\ts+1},\srtime_{\ts+1}]$ and reward $r_\ts = \rew(\state_\ts, \acs_\ts, \state_{\ts+1})$. We define a notation  $\trans_{M,\rtraj}(s'|s,a)$  as the above transition process based on  $M$ and $\rtraj$. Note that instead of directly predicting the whole next state $\state_{\ts+1}$, the simulator just predicts $y$ in the past and constructs the other states from histoical  data $\rtraj$.

The general goal of simulator-based RL is to find an optimal policy $\hat \pi^*$ which maximizes the cumulative reward (i.e., LTE) for all users. In particular, the objective is written as:
\begin{align}
\max_\pi\mathbb{E}_{\group\sim p(\group), \user \sim p(\user), \rtraj \sim \dataset(u,g)}  \left[\mathbb{E}_{\tau \sim p(\tau|\pi,P_{M,\rtraj})} \left[\sum_{\ts=0}^\T \discount^\ts r_{\ts} \right]\right],
\label{eq:sim-rl}
\end{align}
where $p(\group)$ and $p(\user)$ are the prior distributions of groups and users, $ \rtraj \sim \dataset(u,g)$ denotes sampling a real trajectory of user $\user$ in group $\group$ from the logged dataset $\dataset$, and $p(\tau|\pi,P_{M,\rtraj})$ is the probability of generating a trajectory $\tau:=[s_0, a_0, r_0, ..., a_{T-1}, s_T, r_T]$ under the policy $\pi$ and transition function $P_{M,\rtraj}$. In particular,
	\begin{align}
		p(\tau\mid \pi, \trans) := \initdist(s_0) \prod\nolimits_{t=0}^T  \trans(s_{t+1}\mid s_t, a_t) \pi(a_t|s_t)\label{equ:traj-dist},
	\end{align}
where $ \initdist(s_0)$  is the initial state distribution. 

\subsection{Reality-gaps of Simulator-based RL in SRS}

We first define the real user model $\mathcal{E}$ which outputs the real feedback of users. Since a user $\user \in \uspac$ has his/her behavior pattern and also depends on the group $\group \in \gspac$ he/she belongs to, we define two functions $ F_\user(\user)$ and $F_\group(\group)$ to map these individuals to corresponding parameters of behavior patterns. Then, we can construct the real user model as  $\mathcal{E}(y|s,a, F_\user(\user), F_\group(\group))$. The real optimal policy $\pi^*$ is the policy which maximizes:
\begin{align}
	\max_\pi\mathbb{E}_{\group\sim p(\group), \user \sim p(\user)}  \left[\mathbb{E}_{\tau \sim p(\tau|\pi, P_\realworld)} \left[\sum_{\ts=0}^\T \discount^\ts r_{\ts} \right]\right].
	\label{eq:realworld_obj}
\end{align}
Assume that we have correct prior distribution $p(\user)$ and $p(\user)$, which is mild as we can easily control the scope recommended users when deployed. Then, we can see that the reality-gaps come from the mismatching between $P_\realworld$ and $P_{M,\rtraj}$, which makes $\pi^* \neq \hat \pi^*$. The performance gap between $\pi^*$ and $\hat \pi^*$ will be large if the transition gap between  $P_\realworld$ and $P_{M,\rtraj}$ is large~\cite{mopo}.  Moverover,
the one-step prediction error will be compounded in the process of multi-step rollout and finally makes the performance gap larger~\cite{imitation_bound}.

The major notations in this paper are summarized in Table~\ref{table:notation}.
\begin{table}[h]
\vspace{-3mm}
\caption{Major notations.}
\begin{tabular}{c l}
\hline
$\pi$     & \begin{tabular}[c]{@{}l@{}}The recommendation policy \end{tabular} \\ 
$\phi$     & \begin{tabular}[c]{@{}l@{}}The environment-parameter extractor for user model\end{tabular} \\
$z$     & \begin{tabular}[c]{@{}l@{}}The environment parameters inferred with $\phi$\end{tabular} \\
$F_\group$ and $F_\user$     & \begin{tabular}[c]{@{}l@{}}The functions to map group $\group$ and  user $\user$ to \\ corresponding parameters of behavior patterns\end{tabular} \\
$\realworld$     & \begin{tabular}[c]{@{}l@{}}The real user feedback model \end{tabular} \\
$M_\omega$     & \begin{tabular}[c]{@{}l@{}}The user simulator parameterized by $\omega$\end{tabular} \\
$\Omega$     & \begin{tabular}[c]{@{}l@{}}The parameter space of $\omega$\end{tabular} \\
$\alg$     & \begin{tabular}[c]{@{}l@{}}The user-simulator learning algorithm\end{tabular} \\
$X^g_t$     & \begin{tabular}[c]{@{}l@{}}The state-action trajectory of group $g$ before timetep $t$\end{tabular} \\
$\psi^g_t$     & \begin{tabular}[c]{@{}l@{}}The parameters of the distribution which generates the \\ state-action pairs in $X^g_t$\end{tabular} \\
$\theta$     & \begin{tabular}[c]{@{}l@{}}The parameters of the posterior approximation in SADAE\end{tabular} \\
$\kappa$     & The parameters of the inference process in SADAE\\ \hline
\end{tabular}
\label{table:notation}
\end{table}

%% file: method.tex
\vspace{-3mm}
\section{Simulation to Recommendation}
\subsection{Zero-shot Policy Transfer Framework}
\label{sec:formulation}

In this section, we introduce zero-shot policy transfer techniques into SRS. 
Standard zero-shot policy transfer techniques have been widely used to overcome the reality-gaps of physical simulators in challenging tasks~\cite{osi-lstm,OpenAIcube,dr}. 
These techniques assume the reality-gaps come from the gap of environment parameters $\omega$. In general, they first construct a simulator set with a massive number of different environment parameters $\omega$ from the environment-parameter space $\Omega$. Based on the simulator set, they learn an extractor $\phi$ to infer the environment parameters, and a context-aware policy $\pi$ to control an agent to perform adaptable behaviors for optimal performance according to the inferred parameters~\cite{osi-lstm,OpenAIcube}.  When deployed, the extractor adjusts its inferred environment parameters via the real interaction trajectory information and thus adapts the policy to suitable behaviors automatically. 

In SRS, since users' behaviors are often hard to model via physical rules, many practical applications learn to predict the behaviors via data-driven techniques~\cite{usersim,virtualtaobao,demer,debiase-1}.  Here we assume the user simulator $M_\omega$ is parameterized by $\omega$ which is learned through a user-simulator learning algorithm $\alg$. Then $\realworld$, $F_\user$ and $F_\group$ are implicitly represented by $\omega$ based on $\alg$.

Now we adopt the standard zero-shot policy transfer framework into SRS~\cite{OpenAIcube,osi-lstm,osi,dr,dr1,bayrn}. Formally, we propose the following objective to handle the reality-gap problem:
\begin{align*}
	\max_{\pi} \mathbb{E}_{\omega \sim p(\Omega)} \left[\mathbb{E}_{\rtraj \sim p(\rtraj), \tau \sim p(\tau|\pi, \phi, P_{M_\omega,\rtraj})} \left[\sum_{\ts=0}^\T \discount^\ts r_{\ts} \right]\right],
\end{align*}
where $\Omega$ is the parameter space of $M$, $p(\Omega)$ is a sampling strategy for model's parameters generation, $p(\rtraj)$ is a simplification of the process $\rtraj\sim \dataset(\user,\group), \user \sim p(\user), \group \sim p(\group)$ (see Eq.~\eqref{eq:sim-rl}), and $p(\tau|\pi, \phi, P_{M_\omega,\rtraj})$ denotes a rollout process based on a context-aware policy $\pi$ and environment parameter extractor $\phi$: for each time-step $\ts$, \textit{we first infer the environment-parameter of current user model $M_\omega$, $z=\phi(M_\omega)$} (we will discuss the specific input of $\phi$ later), where $z$ is the representation of the user model $M_\omega$, then a context-aware policy $\pi(a|s,z)$ will take actions based \textit{on the inferred representation $z$}.  The context-aware policy $\pi$ is trained to make the optimal decisions in all of the models $M_\omega$ where $\omega \in \Omega$. When deployed, we use the same extractor to infer the representation of the real-world $z_r =\phi(\realworld)$, then the context-aware policy $\phi$ makes decisions based on $z_r$: $a\sim \hat \pi^*(a|s,z_r)$. If the $\realworld$ can be represented by $\omega$, that is, $\exists \omega \in \Omega, M_\omega \approx \realworld$, and $\phi$ can identify the representation of parameters correctly, we have $\hat \pi^*(a|s,z_r)\approx \pi^*(a|s), \forall s \in \sspac$~\cite{OpenAIcube,osi-lstm,osi}. 

However, the above solution is infeasible in practice because of  the following two aspects:

(1) \textbf{extremely large parameter space of $\Omega$}: In previous applications of zero-shot policy transfer techniques~\cite{osi,osi-lstm,OpenAIcube,bayrn,anymal_wild}, $\trans$ is built through physics principles with some parameters $\omega$ with specific definitions, like friction coefficients or lengths of robot arms. Thus the space $\Omega$ is compact for $\phi$ and $\pi$ learning. In SRS, the simulator is built via data-driven techniques, then $\omega$ is complex, e.g., the weights of neural networks. Thus the space of $\Omega$  is large and redundant. Currently, it is almost impractical to develop $\phi$ and $\pi$ to identify $\omega$ from such a large space directly. To develop a practical zero-shot policy transfer technique for SRS, we should shrink $\Omega$ to a feasible scale firstly;

(2) \textbf{the high complexity of $\phi$ to identify correct representations}: In previous applications, the policy is to operate a single robot (like quadruped robots~\cite{anymal_wild}, robot arms~\cite{osi-lstm}, or robot hands~\cite{OpenAIcube}). They only need to identify the parameters of the deployed robot.  Thus it is feasible for some practical online searching methods to search the correct parameters directly via some online interaction samples~\citep{bayrn}. In SRS, the policy serves numerous users in multiple regions at the same time. Thus the computing cost will be large for searching the parameters for all of the users and will be unacceptable in large-scale internet platforms. 
Another paradigm is representation learning: they train an environment-parameter  extractor $\phi$ to embed historical interaction samples of the agent to hidden variables $z$. A recurrent neural network (RNN) is often used to embed the sequential information into environment-parameter vectors $z_t = \phi(s_t, a_{t-1}, z_{t-1})$. In theory, the target environments are identifiable after embedding enough interaction samples. This pipeline is more suitable to SRS scenario as the end-to-end inference module $\phi$ has less computing cost when deployed. However, in SRS, the user's feedback is not only dependent on user's personas ($F_\user$ in Eq.~\eqref{eq:realworld_obj}) but also dependent on user's region ($F_\group$ in Eq.~\eqref{eq:realworld_obj}).  It needs much more time-steps of interactions for identifying $z$ if only considering single-user's interactions, which leads to extra  risks of decision-making when deployed, as the policy needs more steps for probing to identify the optimal policy for each user in each group~\cite{maple}.

As discussed above, representation learning of $\phi$ is a paradigm with potential to handle the reality-gap problem in SRS. In this article, we follow this paradigm and propose several practical techniques to solve the above challenges. Formally, to find the optimal extractor for $\phi^*$ and policy $\pi^*$, a standard objective~\cite{OpenAIcube,osi-lstm} is:
\begin{equation}
	\max_{\pi,\phi} \mathbb{E}_{\omega \sim p(\Omega)} \left[\mathbb{E}_{\rtraj \sim p(\rtraj), \tau \sim p(\tau|\pi, \phi, P_{M_\omega,\rtraj})} \left[\sum_{\ts=0}^\T \discount^\ts r_{\ts} \right]\right],
	\label{overral:obj}
\end{equation}
where $p(\Omega)$ denotes a sample strategy to draw transition functions $M_\omega$ from the simulator parameter set $\Omega$, s.t., $P[\omega]>0, \forall \omega \in \Omega$.  
 We take a uniform sampling strategy in the following analysis. 
For each time-step $\ts$, we first infer the environment-parameter via $z_{\ts}=\phi(\state_\ts, \acs_{\ts-1}, z_{\ts-1})$, where $\state_\ts$ is a sample of $\trans_{M_{\omega}}(\state|\state_{\ts-1},\acs_{\ts-1})$, then a context-aware policy $\pi(a|\state_{\ts},z_{\ts})$ will take actions based on the inferred representation $z_{\ts}$.
$\phi$ and $\pi$ are optimized together via Eq.~\eqref{overral:obj}. \textit{Note that the gradients would be backpropagated from $\pi$ to $z$  if optimal policies in different simulators are inconsistent but have the same representation of $z$, then the parameters of $\phi$ is updated automatically to identify the parameter in $\Omega$.}
  
In the following of the article, we first propose a new environment-parameter extractor architecture of $\phi$ for more efficient parameter identification in the SRS scenario, which is in Sec.~\ref{sec:phi}.  In Sec.~\ref{sec:reliable}, we develop several techniques to reduce $\Omega$ to a feasible scale for $\pi$ and $\phi$ learning.

\subsection{Hierarchical Environment-parameter Extractor}
\label{sec:phi}

In SRS, environment parameters are dependent on user and group information $\user$  and $\group$.  If we have the ground-truth features of $\user$ and $\group$, we can feed them into $\phi$: $z_t = \phi(s_t, a_{t-1}, z_{t-1}, \group, \user)$ to solve the representation identification problem. However, it is inevitable having some features of $\gspac$ and $\uspac$ that are hard to model. 
Therefore, besides constructing static states via feature engineering related to $\mathcal{G}$ and $\mathcal{U}$, i.e., $\suser$ and $\sgroup$, we develop a hierarchical architecture of the extractor for modeling user and group information.
Intuitively, we should add the group trajectory $S^\group_0, A^\group_0, S^\group_1, A^\group_1, ..., S^\group_t$ to the input, that is $z_t = \phi(s_{t}, a_{t-1}, S^g_{t}, A^g_{t-1}, z_{t-1})$, where $(S^g_t, A^g_{t-1}):= \{(s^{(i)}_t, a^{(i)}_{t-1})\}^N_{i=1}$ (in the rest of this article, we use $X^g_t:=(S^g_t, A^g_{t-1})$ for brevity), which includes $N$ state-action pairs at each time-step $t$. 

However, the user number $N$ can be large. It is impractical to feed $X^g_t$ to the neural network directly. We prefer to embed $X^g_t$ to a low-dimensional vector $\upsilon$ to feed into $\phi$. Calculating the statistics of $X^g_t$ (e.g., mean and standard deviation) is a direct way but limits the representation capacity of $\upsilon$.
Popular modules like Attention~\cite{attention} are potential, but these modules are computation costly.

\begin{figure}[h]
	\centering
	\includegraphics[width=0.8\linewidth]{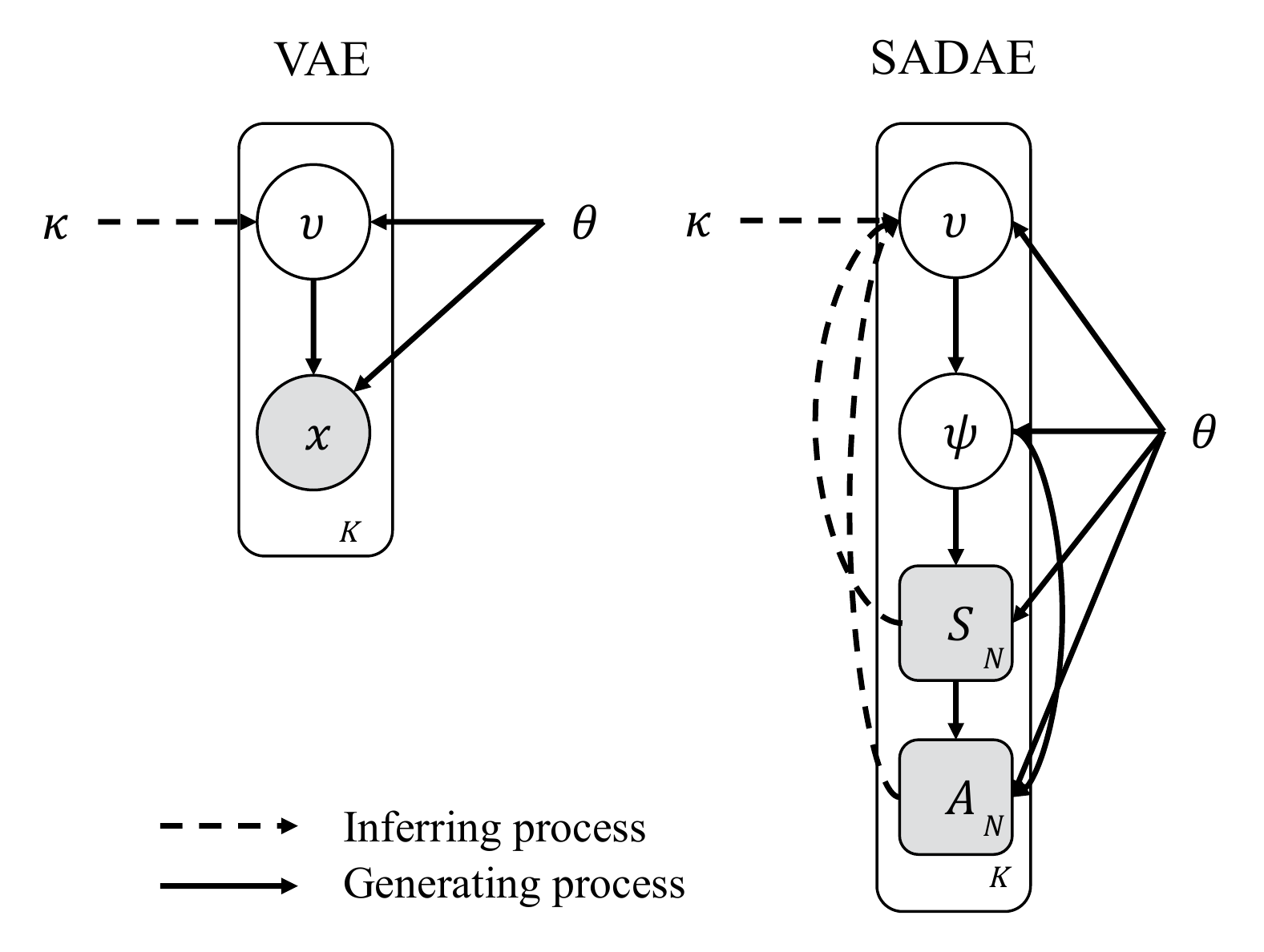}
	\caption{Comparison of SADAE and vanilla VAE through the directed graphical model. The circles denote the variable nodes. The rounded rectangle denotes the dataset nodes, in which the notation in the corner denotes the number of datasets. $\theta$ denotes the approximation parameters of the generative model, $\kappa$ denotes the parameters of the variational approximation model, $K$ denotes the number of samples of $S$ and $A$ in $\dataset$, and $N$ denotes the number of samples of $s$ and $a$ in $S$ and $A$.
	}
	\label{fig:SADAE}
\end{figure}

In this work, we propose a simple way to infer a latent embedding $\upsilon$ given $X^g_t$, named \textbf{S}tate-\textbf{A}ction \textbf{D}istributional variational \textbf{A}uto\textbf{E}ncoder (SADAE) inspired by variational autoencoder (VAE)~\cite{KingmaW13}.  We first formulate the data generative process based on the assumptions: First, state-action pairs in $X^g_t$ are i.i.d. sampled from a distribution $p_{\psi^g_t}(s, a)$ parameterized by $\psi^g_t$ for each time-step $t$ and group $g$.  Second, the parameters $\psi$ of the distribution are generated by a distribution $p_\theta(\psi\mid\upsilon)$ parameterized by $\theta$. It involves a latent continuous random variable $\upsilon$, which is generated from a prior distribution $p(\upsilon)$.  The generation of $X$ includes three steps: (1)  sample $\upsilon$ from $p(\upsilon)$; (2) sample $\psi$ from distribution $p_\theta(\psi\mid\upsilon)$ ; (3) sample $p_\psi(s, a)$ repeatedly to generate $X$. 
A comparison with VAE on directed graphical model is shown in Fig.~\ref{fig:SADAE}.
	
Formally, our target is to learn an embedding model $q_\kappa(\upsilon\mid X)$ parameterized by $\kappa$, aligned with the posterior approximation  $p_\theta(\upsilon\mid X)$. Using Kullback-Leibler Divergence (KLD) as the measurement, the objective can be written as follows: 
	\begin{align}\label{equ:kld-main}
	\min_{\kappa,\theta} \mathbb{E}_{X\sim\dataset} \left[KLD\left(q_\kappa(\upsilon\mid X)\mid \mid p_\theta(\upsilon\mid X)\right) \right], 
	\end{align}
where the dataset $\dataset$ is reshaped to  $\{X^g_t: g \in \mathcal{G}, 0<t\le T \}$ includes state-action pairs in all time-steps and groups, and the posterior $p_\theta(\upsilon\mid X)$ is the target distribution of $q_\kappa(\upsilon\mid X)$.  For brevity, we use $\theta$ and $\kappa$ to denote all parameters of posterior approximation and inference respectively.

We first provide the evidence lower bound (ELBO) of Eq.~\eqref{equ:kld-main} in Lemma~\ref{lemma:elbo}.
\begin{lemma} The ELBO of state-action distributional variational inference objective Eq.~\eqref{equ:kld-main} is:
	\label{lemma:elbo}
	\begin{small}
	\begin{equation*}
	\max_{\kappa,\theta}	\mathbb{E}_{X\sim\dataset}\left[ \mathbb{E}_{q_\kappa(\upsilon\mid X)}\left[\log p_\theta(X\mid \upsilon) \right] - KLD\left(q_\kappa(\upsilon\mid X) \Vert p_\theta(\upsilon) \right) \right]. \label{elbo}
	\end{equation*}
\end{small}
\end{lemma}

Under the assumption of i.i.d. on $X$, $q_\kappa(\upsilon\mid X)$ and $p_\theta(X\mid \upsilon)$ can be estimated via likelihood: 
\begin{small}
	\begin{align}
	q_\kappa(\upsilon\mid X) &= \prod_{i=1}^N q_\kappa\left(\upsilon \mid s^{(i)} a^{(i)}\right), \label{equ:infer-main} \\ 
	p_\theta(X \mid \upsilon) 
	&= \prod_{i=1}^N p_\psi \left(s^{(i)}, a^{(i)} \right)  p_\theta(\psi\mid \upsilon),  \label{equ:recons-main}
	\end{align}
\end{small}
	where $\psi$ denotes the parameters of distribution $p_\psi$. We give our theorem of the tractable evidence lower bound (ELBO) in Theorem~\ref{the:1-main}. We leave the proof in Appendix.~\ref{app:proof}.

	\begin{theorem}\label{the:1-main} The tractable ELBO of state-action distributional variational inference is:
	\begin{small}
	\begin{align}
	    &  \mathbb{E}_{X\sim\dataset,q_\kappa(\upsilon\mid X)}\Big[ \sum_{i=1}^N  \log p_\theta\left(s^{(i)}\mid \upsilon \right) \Big. 
		+  \Big. \log p_\theta\left(a^{(i)}\mid \upsilon, s^{(i)}\right) \Big]  \nonumber\\ 
		- & \Bigg.  KLD\left(  q_\kappa(\upsilon\mid X) \mid \mid p_\theta(\upsilon) \right).
	\label{sadae-loss}
	\end{align}
	\end{small}
	\end{theorem}

\begin{figure}
	\centering
	\includegraphics[width=1.0\linewidth]{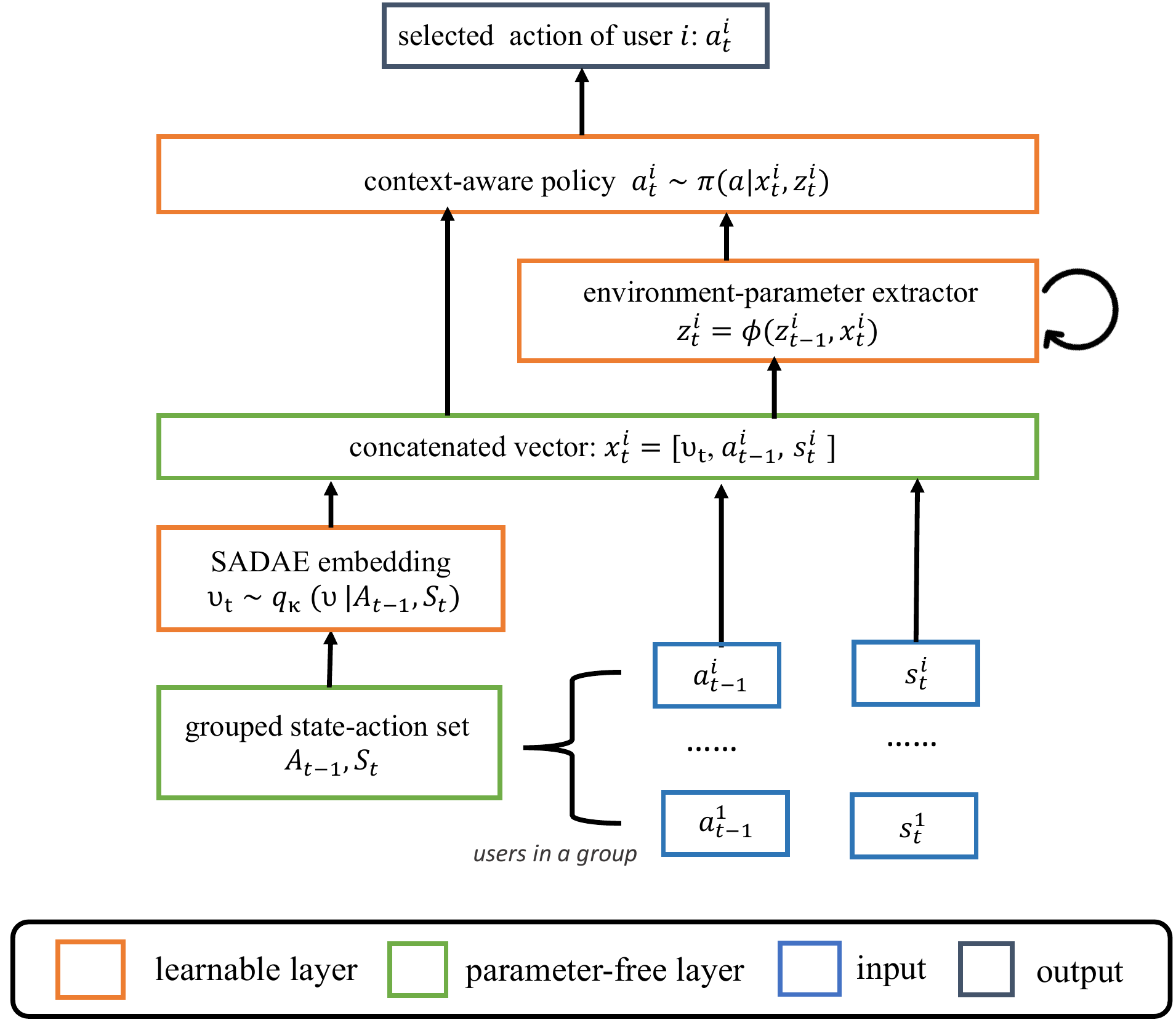}
	\caption{The overall architecture of Sim2Rec.}
	\label{fig:struc}
	\vspace{-5mm}
\end{figure}

Theorem~\ref{the:1-main} gives us a three-step pipeline to minimize the objective of~Eq.~\eqref{equ:kld-main}: (1) sample a batch of $X$ from the dataset $\dataset$; (2) infer latent code $\upsilon$ via Eq.~\eqref{equ:infer-main}; (3) compute the reconstructed log-probability of state-action pairs based on Eq.~\eqref{equ:recons-main} and KL divergence between posterior and prior of $\upsilon$, and then apply the  gradient to $\kappa$ and $\theta$. Finally, the extractor $\phi$ infers environment-parameter both with $s$, $a$ and $\upsilon$: $z_t = \phi(s_t, a_{t-1}, \upsilon_t, z_{t-1})$, where $\upsilon_t \sim q_{\kappa}(\upsilon \mid X_t)$. Then the context-aware policy $\pi\left(a_t \mid s_t, z_t \right)$ samples an action based on $z_t$. $q_{\kappa}$ is also updated with Eq.~\eqref{overral:obj}. The gradient will be backpropagated from $\phi$ to $\upsilon$ to update $\kappa$. The overall architecture is shown in Fig.~\ref{fig:struc}.

    \subsection{Feasible Parameter Space $\Omega$ Construction}\label{sec:reliable}
    
    Considering a data-driven user simulator based on neural networks, the original parameter space of $\Omega$  will be the weight space of the neural networks, which will be extremely large and complex. 
    
    However, many of these weights $\omega \in \Omega$ cannot imitate the feedback of the users at all.  It is unnecessary to make $\phi$ and $\pi$ to be aware of all of the weights in $\Omega$. In this perspective, user simulator imitation algorithms $\alg$~\cite{usersim,demer,virtualtaobao} can be regarded as a practical way to sample a $\omega$ which is close to the real-world's parameters $\omega^*$. Specifically, we have $\omega =\alg(\dataset, \lambda)$, where $\dataset$ is the dataset for user simulator learning and $\lambda$ is the hyper-parameters (e.g., random seeds and learning rates) of the learning algorithm $\alg$. With different $\dataset$ and  $\lambda$, $\alg$ will generate a weight vector $\omega$. 
    
    Inspired by ensemble techniques, in this work, we construct a shrunken parameter space $\Omega':=\{\omega: \alg(\dataset', \lambda), \lambda \in \Lambda, \dataset' \subseteq \dataset\}$, where  $\dataset'$ is a subset of $\dataset$ and $\Lambda$ is the selected hyper-parameters space for  $\alg$ learning. In this way, we can generate a weight set where $\omega \in \Omega'$ are roughly close to $\omega^*$ with suitable  $\dataset'$ and  $\lambda$.
    
    However, we still cannot claim that $\omega^* \in \Omega'$, since the user behavior is too complex to be predicted exactly. All of $\omega$ might have prediction errors in some states and actions. 
   In general, the learning errors include the two aspects: (1) The approximation error: the approximation error in the dataset $\mathcal{D}$ is limited by the capacity of the neural network models and learning tools; Besides, in sequential environments, the approximation error will be inevitably compounded for each step, leading to a large discrepancy of simulation trajectories even if the one-step prediction error is small~\cite{mopo}; (2) The extrapolation error: since these models are used as a simulator for policy training, we expect the models to give unbiased predictions when querying with other actions except for the data-collection actions. This makes the model learning and using in the data distributions violate the independent and identically distributed (\emph{i.i.d.}) assumption and leads to the extrapolation errors. Then the predictions might be catastrophic failures in unseen actions, a.k.a. counterfactual actions, and will be totally wrong in guiding policy learning~\cite{galileo,offline-overview}.
    
    Although we cannot make $\omega \in \Omega'$ hold directly, we can intervene the exploration process of RL to avoid the policy learning in regions where the gaps between $M_{\omega^*}$ and $M_{\omega}$ are large. 
For generality, in this article, we design several post-processing methods agnostic to specific model learning techniques to handle the above problems:

\begin{algorithm}[h]
	\caption{Sim2Rec pseudocode}
	\label{alg:sim2rec}
	\begin{flushleft}
		\textbf{Input}:\\
		$\phi_\varphi$ as an environment-parameter extractor, parameterized by $\varphi$; 
		Context-aware policy $\pi_\iota$ parameterized by $\iota$; state-action distributional embedding $q_\kappa$;
		Logged dataset $\dataset$; 
        coefficient of uncertainty penalty: $\alpha$; 	Truncated rollout horizon $T_c$; Model uncertainty function $U$;\\
		
		\textbf{Process}:
	\end{flushleft}	
	\begin{algorithmic}[1]
		 \State Construct the parameter set $\Omega':=\{\omega: \alg(\dataset', \lambda), \lambda \in \Lambda, \dataset' \subseteq \dataset\}$; \label{alg:Sim2Rec:demer}
		 \State Initialize an empty buffer $\mathcal{D}_\text{rollout}$; 
		\For{1, 2, 3, ...}  
		\State Sample a simulator $M_\omega$, where $\omega \sim p(\Omega)$.
		\State Select a group $g \sim p(g)$.
		\State Sample real trajectories $\tau^r \sim p(\tau^R)$ from the group $g$ and sample simulation trajectories $\tau \sim p(\tau|\pi, \phi, P_{M_\omega,\rtraj})$ with the truncated horizon $T_c$. \label{alg:collect}
		\State Add the trajectories $\tau$ to $\dataset_{\rm rollout}$.
		\State Add uncertainty penalty $U(s,a)$ to the rewards in $\dataset_{\rm rollout}$, i.e., $r \leftarrow r -\alpha U(s,a)$.
		\State Filter data in $ \dataset_{\rm rollout}$ through $\mathcal{D}_{\rm rollout} \leftarrow F_{\rm trend}(\dataset_{\rm rollout})$ and update the done and reward value through $\mathcal{D}_{\rm rollout} \leftarrow F_{\rm exec}(\dataset_{\rm rollout})$.
			\State Update $\varphi$, $\theta$, $\iota$ and $\kappa$ via Eq.~\eqref{overral:obj}  with $\mathcal{D}_\text{rollout}$ using one of RL algorithm and update $\kappa$ via Eq.~\eqref{sadae-loss}.
		\EndFor
	\end{algorithmic}
\end{algorithm}

   \textbf{Avoid the policy exploiting the regions with large prediction errors}:
   To avoid the agent reaching regions that might be given wrong predictions with high probability, at each step, a penalty is added to the reward which is calculated according to the model uncertainty $U(s_t,a_t)$~\cite{mopo}. 
   The model uncertainty $U$ measures the inconsistency of prediction among the learned transition models at $(s_t, a_t)$;
   To mitigate the compounding error of the models, we randomly draw a state from the logged dataset as the initial state and constrain the maximum rollout length to a fixed number $T_c$. The above solutions are inspired by~\cite{mopo,maple}, which are offline model-based RL algorithms in MuJoCo~\cite{mujoco}.
   
   \textbf{Guarantee the policy optimizing in the regions without large extrapolation errors:} 
   In the LTE optimization problem, we often have prior knowledge on the trend of user feedback for the changing of action given a specific application. For example, for demand prediction, if the price is increased, the demand of users would be decreased. 
   Taking use of the prior knowledge of elasticity, we can evaluate the prediction of models to counterfactual actions and remove the trajectories in $\mathcal{D}$ where the predictions of user simulator $M$ is inconsistent with the prior of the tendency. 
  We use $\mathcal{D} \leftarrow F_{\rm trend}(\mathcal{D})$ to denote the filter process.
  Besides, we define the executable action subspace for each state to avoid policy taking actions far away from the data-collection policy $\pi_e$. For example, in our application, we calculate the minimal and maximal action values $a_{\rm min}^\user, a_{\rm max}^\user$ that have ever been taken by $\pi_e$ in historical interactions for user $\user$. 
   If the output of policy $a \notin (a_{\rm min}^\user, a_{\rm max}^\user)$, to avoid the policy taking the risky action, the state can be set to a done state, i.e., $ {\rm done}=\mathbb{I}\left[a \notin \left(a_{\rm min}^\user, a_{\rm max}^\user\right)\right]$,  and the reward can be set to $\frac{R_{\rm min}}{1-\gamma}$ where $R_{\rm min}$ is the minimal reward of the task. We use $\mathcal{D} \leftarrow F_{\rm exec}(\mathcal{D})$ to denote the process.

Based on the above techniques, we give the pseudocode of \our{} in Alg.~\ref{alg:sim2rec}.

%% file: experiment.tex
\section{Experiments} \label{sec:exp}

In this section, we first conduct experiments~\footnote{We release our code at \url{https://github.com/xionghuichen/Sim2Rec}} in a synthetic recommendation environment in Google RecSim~\cite{RecSim}, named the long-term satisfaction (LTS). We then apply \our{} to the driver program recommendation (DPR) task in a large-scale ride-hailing platform, \didi{}, to demonstrate the effectiveness of the proposed method in the real-world setting. In particular,  we mainly focus on the following questions:

\begin{itemize}
    \item \textbf{RQ1:} Whether SADAE can effectively reconstruct the group information?
    \item \textbf{RQ2:}  In the synthetic environment which has predefined feasible environment-parameter space, whether the extractor architecture proposed in Sec.~\ref{sec:phi} can identify the environment more efficiently?
    \item \textbf{RQ3:} Whether the proposed techniques of constructing a feasible parameter space for data-driven simulators in Sec.~\ref{sec:reliable} are useful in real-world applications?
    \item \textbf{RQ4:} Whether the \our{} policy can achieve better performance in unseen environments than the benchmark recommendation systems in real-data tasks?
    \item \textbf{RQ5:} How the whole system performs in a large-scale production environment?
\end{itemize}

In the following, we answer \textbf{RQ1} in Sec.~\ref{exp:sadae-syn} and Sec.~\ref{exp:sadae-real}, \textbf{RQ2} in Sec.~\ref{exp:policy-syn}, \textbf{RQ3} in Sec.~\ref{exp:param-space}, \textbf{RQ4} in Sec.~\ref{exp:policy-real}, and \textbf{RQ5} in Sec.~\ref{exp:ab}. 

In addition, we conduct the ablation studies to validate  the necessity of SADAE proposed in Sec.~\ref{sec:phi} and post-processing methods proposed in Sec.~\ref{sec:reliable}. The experiment results regarding the SADAE and  post-processing methods are analyzed in Sec~\ref{exp:policy-syn} and Sec~\ref{exp:param-space}, respectively.

\subsection{Experimental Setup}
\subsubsection{Implementation Details} 
We use Proximal Policy Optimization (PPO)~\cite{ppo} as the policy learning method to optimize Eq.~\eqref{overral:obj}. The environment-context extractor layer is modeled with a single-layer LSTM network~\cite{lstm}. We add extra fully-connection layers $f$ between the embedding of SADAE $q_\kappa$ and the environment-parameter extractor $\phi$. 
We use the same network structure and hyper-parameters in the two experiments, but the complexity of the neural networks is different. Tab.~\ref{tab:hyper} reports the hyper-parameters.

\begin{table}[h]
	\caption{The hyper-parameters of \our{}. }
	\label{tab:hyper}
	\centering
	\begin{tabular}{lcc}
		\toprule
		Hyperparameter & LTS & DPR\\
		\midrule
    \multicolumn{3}{l}{Policy and extractor learning}\\
    \midrule
		Learning rate & \multicolumn{2}{c}{from 1e-4 to 1e-6}\\
		Optimizer & \multicolumn{2}{c}{Adam}\\
		Discount factor $\discount$& 0.99 & 0.9\\
		Horizon $\T$ & 140 & 30\\
		Batch size & 30000 & 120000\\
		Extra fully-connection layers  $f$ & [128, 128, 128,32] & [512, 512, 256]  \\
		Unit of LSTM in $\phi$ & 64 & 256  \\
		Context-aware layer $\pi$ &[128, 64]& [512, 256]\\
		\midrule
		\midrule
			\multicolumn{3}{l}{SADAE learning}\\
		\midrule
		Embedding layer $q_\kappa(\upsilon|s,a)$ &\multicolumn{2}{c}{[512, 512]}\\
		Reconstructed layer $p_\theta(\psi|\upsilon)$ &\multicolumn{2}{c}{[512, 512]}\\
		Optimizer &\multicolumn{2}{c}{Adam}\\
		Learning rate &2e-5 & 1e-6\\
		L2 regularization weight& 0.1 & 0.001\\
        units of latent code& 5 & 200\\
		\bottomrule
	\end{tabular}
\end{table}

\subsubsection{Baselines} We compare our method \our{} with the following baseline methods:
\begin{itemize}
    \item \textbf{DR-OSI}: An OSI algorithm which uses a standard LSTM neural network as environment-parameter extractor for zero-shot policy transfer~\citep{OpenAIcube}; Compared with \our{}, DR-OSI does not adopt the SADAE for extractor learning in the neural network architecture.
    \item \textbf{DR-UNI}: The domain randomization technique to learn a unified policy~\citep{dr1}. It is an alternative zero-shot policy transfer method which learns a conservative policy from the simulator set. DR-UNI can be regarded as a policy learning method with the same objective as Eq.~\ref{overral:obj} but the output of $\phi$ is a constant. 
    \item  \textbf{DIRECT}: A standard simulator-based policy learning method without considering the reality-gaps of the simulator~\cite{demer};
    \item \textbf{WideDeep}: A supervised learning model for recommendation systems which utilizes wide and deep layers to balance both memorization and generalization~\citep{widedeep};
    \item \textbf{DeepFM}: Also a recommendation systems learning algorithm with a supervised learning method which introduces a factorization-machine layer to replace the wide part WideDeep~\citep{widedeep} and employs deep neural networks to build hybrid structures that exploit the merits of low-order and high-order feature interactions~\citep{deepfm};
    \item \textbf{\our{}-PE}: The \our{} algorithm without using the techniques to handle the prediction errors;
    \item \textbf{\our{}-EE}: The \our{} algorithm without using the techniques, including the two filters $F_{\rm trend}$ and $F_{\rm exex}$, to handle the extrapolation errors.

\end{itemize}

\subsubsection{Evaluation Metrics}
We use  KL divergence to evaluate the distance between the reconstruction data distribution of SADAE and the distribution of the real data, and use the standard metric, long-term rewards, to evaluate the performance of the learned policy. 

\textbf{KL divergence (KLD)}: Since the dimension of state-action space is high and the distribution is complex especially in DPR tasks, we use Kernel Density Estimation (KDE) \cite{kde} to estimate the probability density function (PDF) of real and reconstructed data. Then the KLD of two datasets is computed based on it. In particular, 
\begin{equation}
KLD(\mathcal{D}_a, \mathcal{D}_b) = \frac{1}{||\mathcal{D}_a||} \sum_{x \in \mathcal{D}_a} \log \frac{f_a(x)}{f_b(x)}, \label{equ:kld-compute}
\end{equation}
where $||\mathcal{D}_a||$ denotes the number of samples in the dataset, and $f_a$ and $f_b$ denote the PDF of real and reconstructed data estimated by KDE.

\textbf{Rewards}: The long-term rewards is computed as Eq.~\ref{eq:realworld_obj}. In the LTS task, we sample 750 users for each group for long-term rewards computation. In the DPR tasks, we select all of the drivers for each group for long-term rewards computation.

\subsection{Experiments in the Synthetic Environment}
For better quantify the adaptability of \our{}, we first conduct the experiments in a synthetic LTS simulator in which the environment parameters $\omega$ are configurable~\citep{RecSim}.
\subsubsection{Synthetic Simulator} The long-term satisfaction (Choc/Kale) problem comes from a synthetic environment in the Google RecSim framework~\cite{RecSim}. In this environment, the recommender system sends items of content to users, and the goal is to maximize users' engagement in multiple timesteps. The items of content are characterized by the score of clickbaitiness. The engagement of users is determined by the clickbaitiness score of content and the long-term satisfaction score. 
The higher clickbaitiness score leads to a larger engagement directly but leads to a decrease in long-term satisfaction while the lower clickbaitiness score increases satisfaction but leads to a smaller engagement directly. Moreover, long-term satisfaction is a coefficient to rescale the engagement of the given item of content. 

Formally, the value of engagement for user $i$ at time-step $t$ is sampled from a Gaussian distribution $\mathcal{N}\left(\mu^i_t, {\sigma^i_t}^2\right)$,  which is parameterized by $\mu^i_t:= \left(a^i_t \mu^i_c + \left(1 - a^i_t\right) \mu^i_k \right)  SAT^i_t$ and $\sigma^i_t:=  \left(a^i_t \sigma^i_c + \left(1-a^i_t \right) \sigma^i_k \right)$, 
where $i$ denotes the index of the user, $a^i_t$ denotes the clickbaitiness score of the document item to be recommended. $\mu^i_c$, $\mu^i_k$, $\sigma^i_c$ and $\sigma^i_k$ are hidden states of the user $i$. $SAT^i_t$ denotes the long-term satisfaction score, which is updated by $a_t$: 
\begin{align*}
SAT^i_t&:= {\rm sigmoid}\left(h^i_s \times NPE^i_t \right)\\
NPE^i_t&:= \gamma^i_n NPE^i_{t-1}  - 2 \left(a^i_t - 0.5\right),
\end{align*}
where $NPE^i_t$ denotes the net positive exposure score of the user $i$, $\gamma^i_n$ denotes the memory discount of  $NPE^i_t$, and $h^i_s$ denotes the sensitivity ratio of $NPE$ to satisfaction. $\gamma^i_n$ and $h^i_s$ are also states in this environment. The states $\mu^i_c$, $\mu^i_k$, $\sigma^i_c$, $\sigma^i_k$, $h^i_s$ and $\gamma^i_n$ define the environment parameter. To construct an environment with the multiple groups multiple users, we select $\mu^\user_c$ as the group feature $g$, which are the same among users in a simulator. That is, $\mu_c^i=\mu_c$ for all users $i$.
$\sigma^i_c$, $\sigma^i_k$, $h^i_s$ and $\gamma^i_n$ are the user feature. In particular, the user feature $\user=[\sigma_c, \sigma_k, h_s, \gamma_n, \mu_k]$ and the group feature $g = [\mu_c]$. We randomly sample $h^i_s$ and $\gamma^i_n$ from an uniform distribution for each user at initialization and keep $\sigma^i_c$, $\mu^i_k$ and $\sigma^i_k$ the same among the users and groups. 
However, the observed state $s$ of each user only include ${\rm SAT}^i_{t}$, and $o_i \sim \mathcal{N}(\mu_c, 4)$, and the observed user feedback $y$ is defined as ${\rm SAT}^i_{t+1}$. We use $\realworld(y|s,a,\user,\group)$ to denotes the above process.
We define the parameter space $\omega:=[\omega_\user,\omega_\group]$, $\mu_{c,r}=14$, $\mu_{k,r}=4$, and two mapping function $F_{\omega_\user}(\user)=[\sigma_c, \sigma_k, h_s, \gamma_n, \mu_{k,r}+\omega_\user]$ and $F_{\omega_\group}(\group)=[\mu_{c,r}+\omega_\group]$. Then we can define a user simulator $M_{\omega}(y|s,a):=\realworld(y|s,a,F_{\omega_\user}(\user), F_{\omega_\group}(\group))$ and let $\omega^*:=[0,0]$ as the ``real'' environment to deploy.

Now we can construct the training simulator set by selecting $\omega_g$ directly and control the difference of $\omega_g$ between the training set and the target environment to design different tasks. 
In particular, we construct the target simulator with $\omega^*$ and select the training simulator set by equidistant sampling parameters $\omega_g$ from the space and remove those $\mid \omega_g-\omega^*\mid < \alpha$. $\omega_g$ controls the group behavior.  With larger $\alpha$, the group-behavior difference between the training set and the target environment is larger. In particular, we construct the following tasks:

\begin{itemize}
	\item LTS1: $\Omega=\{\omega: \mid \omega_g \mid \ge 2 \wedge 6\le \mu_c + \omega_g  < 22,  \omega_g \in \mathbb{N}, \omega_\user=0\}$;
	\item LTS2: $\Omega=\{\omega: \mid \omega_g \mid \ge 3 \wedge 6\le \mu_c + \omega_g  < 22,  \omega_g \in \mathbb{N}, \omega_\user=0\}$;
	\item LTS3: $\Omega=\{\omega: \mid \omega_g \mid \ge 4 \wedge 6\le \mu_c + \omega_g  < 22,  \omega_g \in \mathbb{N}, \omega_\user=0\}$;
	\item LTS3-$\beta$: $\Omega=\{\omega: \mid \omega_g \mid \ge 4 \wedge 6\le \mu_c + \omega_g  < 22,  \omega_g \in \mathbb{N}, \omega_\user \in [-\beta,\beta]\}$;	
\end{itemize}
where $\sigma^i_c=\sigma^i_k=1$ for all of the tasks. For simplification, in LTS1 to LTS3, we only consider the reality-gaps of $\omega_\group$.

\subsubsection{Implementations}
In the LTS environment, $\group$ is only related to group state information $S$. Thus we train SADAE to reconstruct the state distribution instead of the state-action distribution. We draw 1000 users for each simulator in LTS3 to the constructed state dataset $\dataset$. $q_\kappa\left(\upsilon|s^{(i)}\right)$ is a neural network which outputs the Gaussian distribution parameters of $\upsilon$.  We also model $p_\theta\left(\psi_s|\upsilon\right)$ with a neural network, which outputs the parameters of Gaussian distributions. The prior of $\upsilon$ is set to standard normal distribution, i.e., $p(\upsilon)= \mathcal{N}(0,1)$.

\subsubsection{Results of Group Information Reconstruction (\textbf{RQ1})}
\label{exp:sadae-syn}

    \begin{figure}[ht]
    	\centering
    	\includegraphics[width=0.6\linewidth]{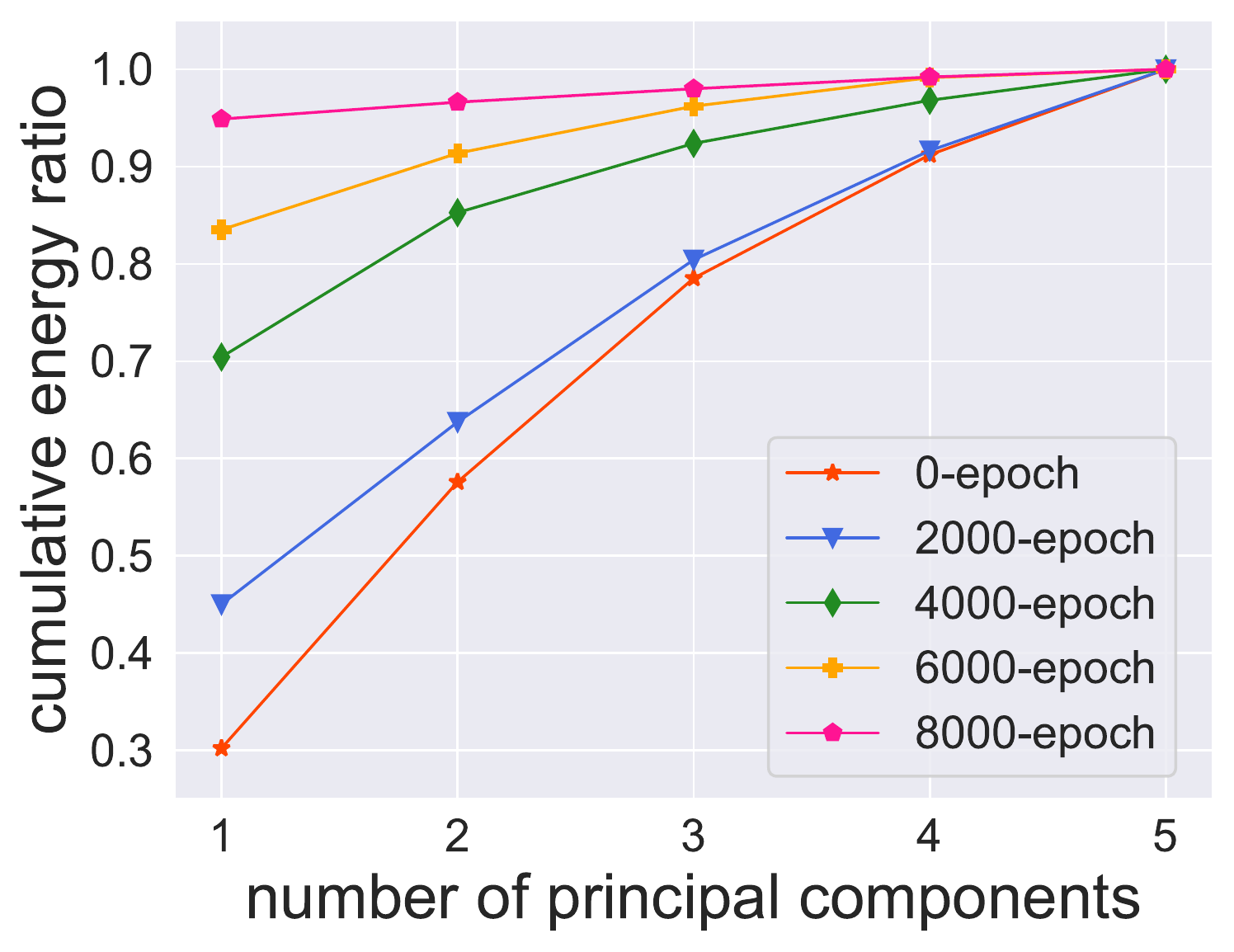}
	    \vspace{-2mm}
    	\caption{Illustration of the cumulative energy ratio with respect to the number of $\upsilon$'s  principal components.
	The energy is represented by the eigenvalue of $\upsilon$'s covariance matrix. The X-axis denotes the number of principal components, and the Y-axis denotes the cumulative energy ratio of principal components. The visualization of projecting $\upsilon$ into two-dimensional vectors based on the first two principal components is in Appendix~\ref{app:pca}. 
	}
	\label{fig:lts:pca}
	    \vspace{-2mm}
    \end{figure}

\begin{figure}[h]
	\centering
	\subfigure[KLD in the training set]{
		\includegraphics[width=0.45\linewidth]{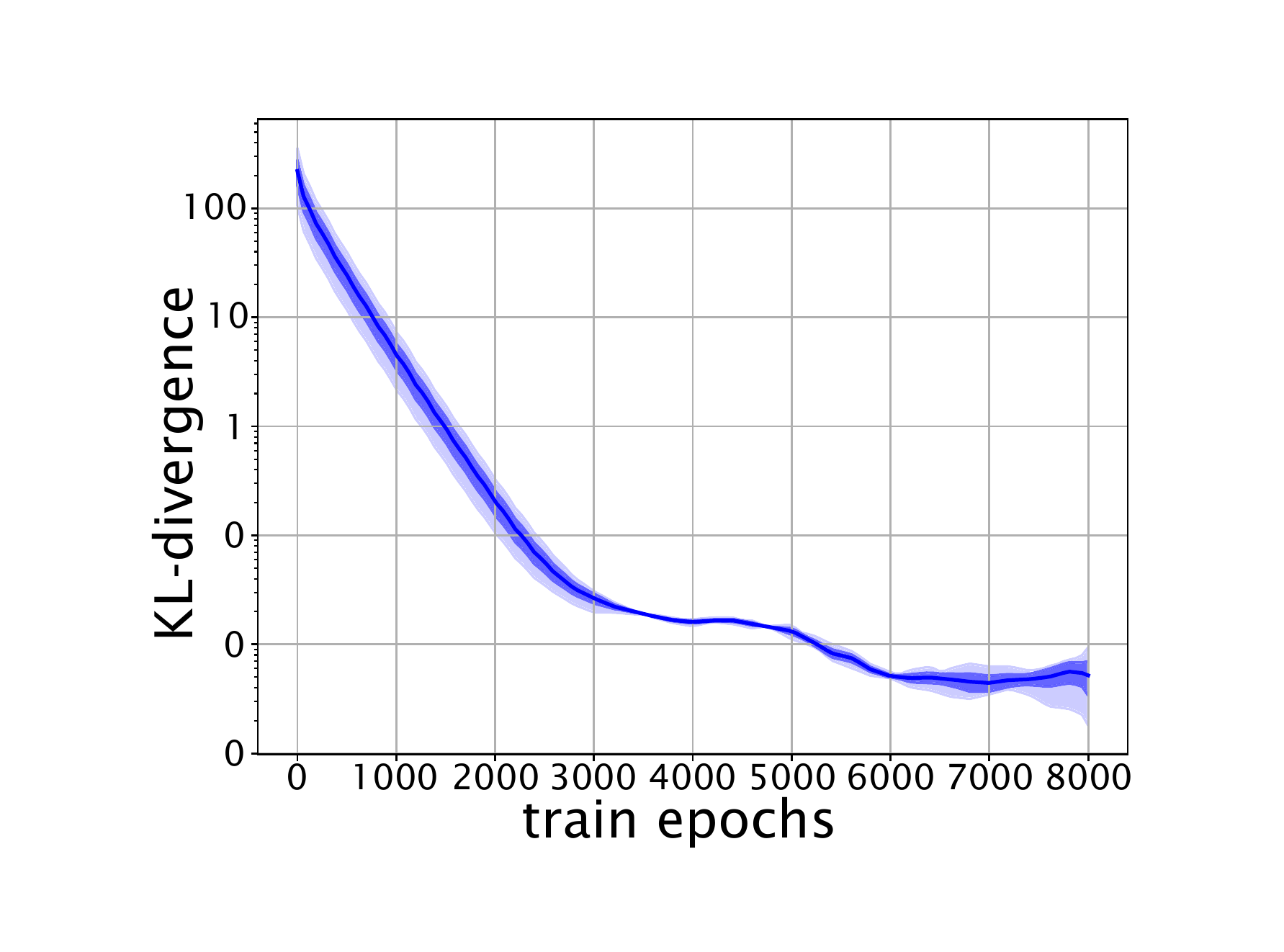}
	}
	\subfigure[KLD in the testing set]{
		\includegraphics[width=0.45\linewidth]{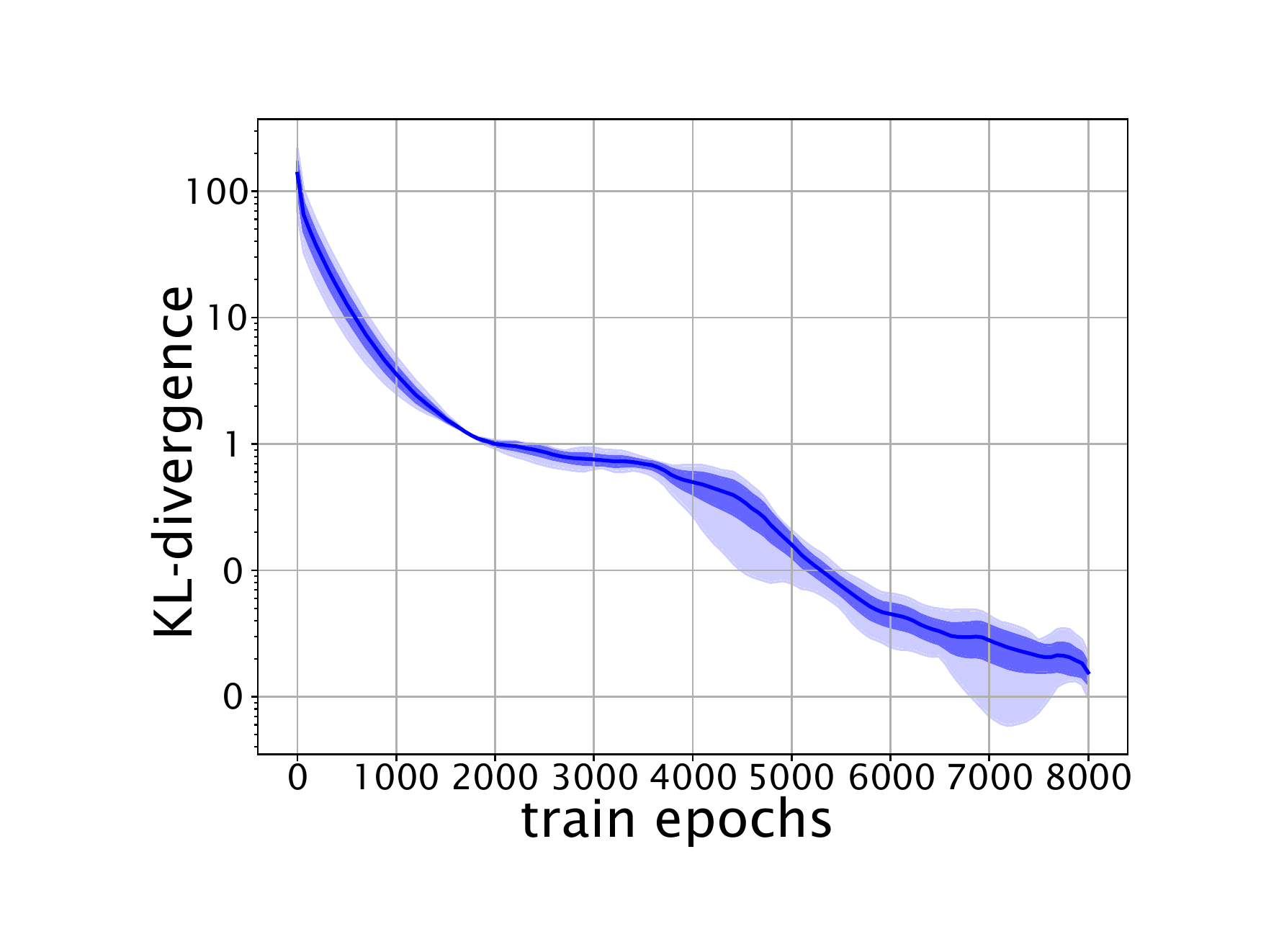}
	}
	
	\caption{Illustration of KL divergence of the training set and testing set in LTS3. The solid curves are the mean reward of three seeds. The dark shadow is the standard error, and the light one is the min-max range of three seeds.}
	\label{fig:lts:vae-kld}
\end{figure}

\begin{figure}[h]
	\centering
		\subfigure[histogram in the traning set ($\mu_c=6$) at epoch 0]{
		\includegraphics[width=0.45\linewidth]{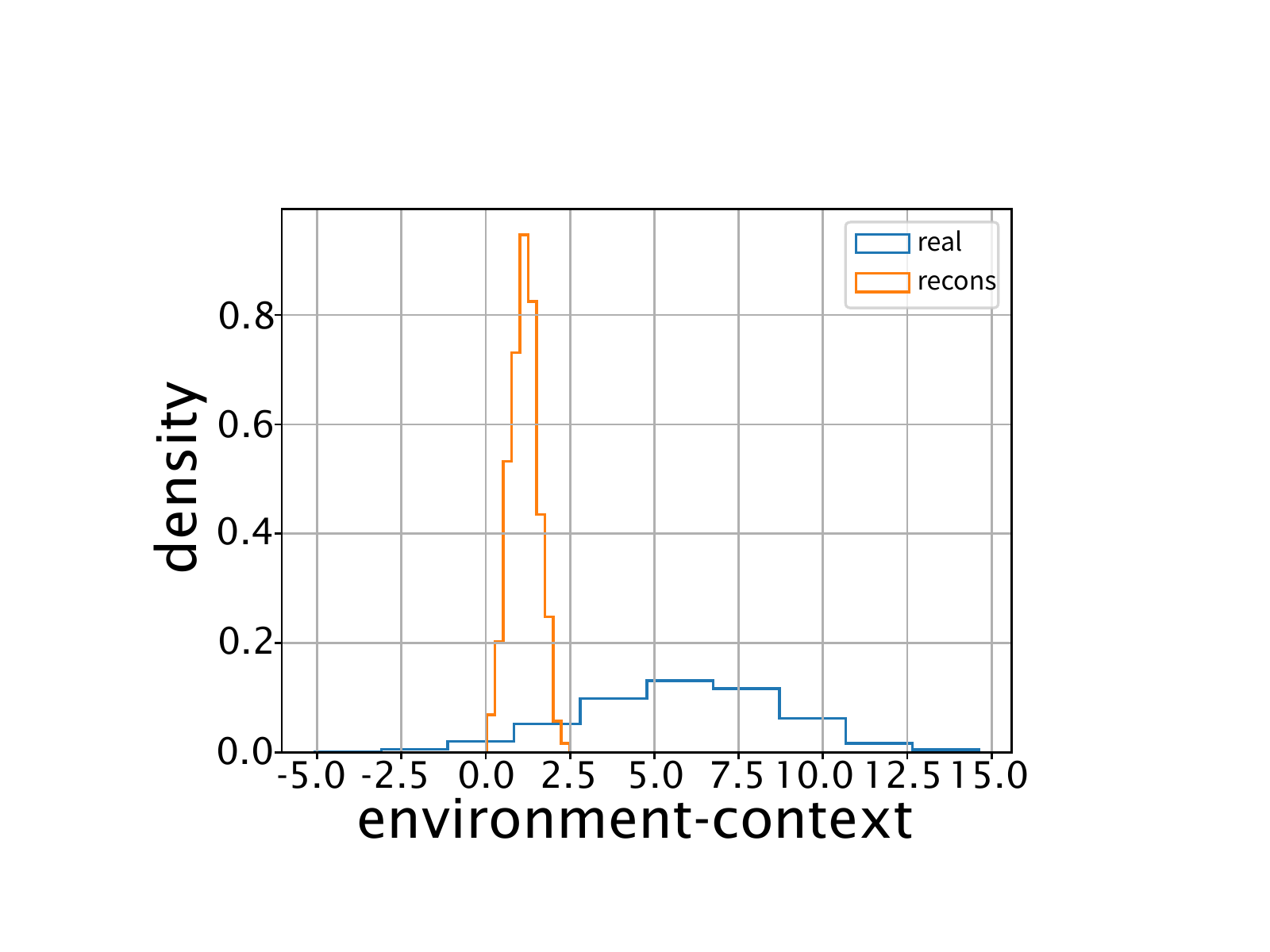}
	}
	\subfigure[histogram in the testing set  ($\mu_c=14$) at epoch 0]{
		\includegraphics[width=0.45\linewidth]{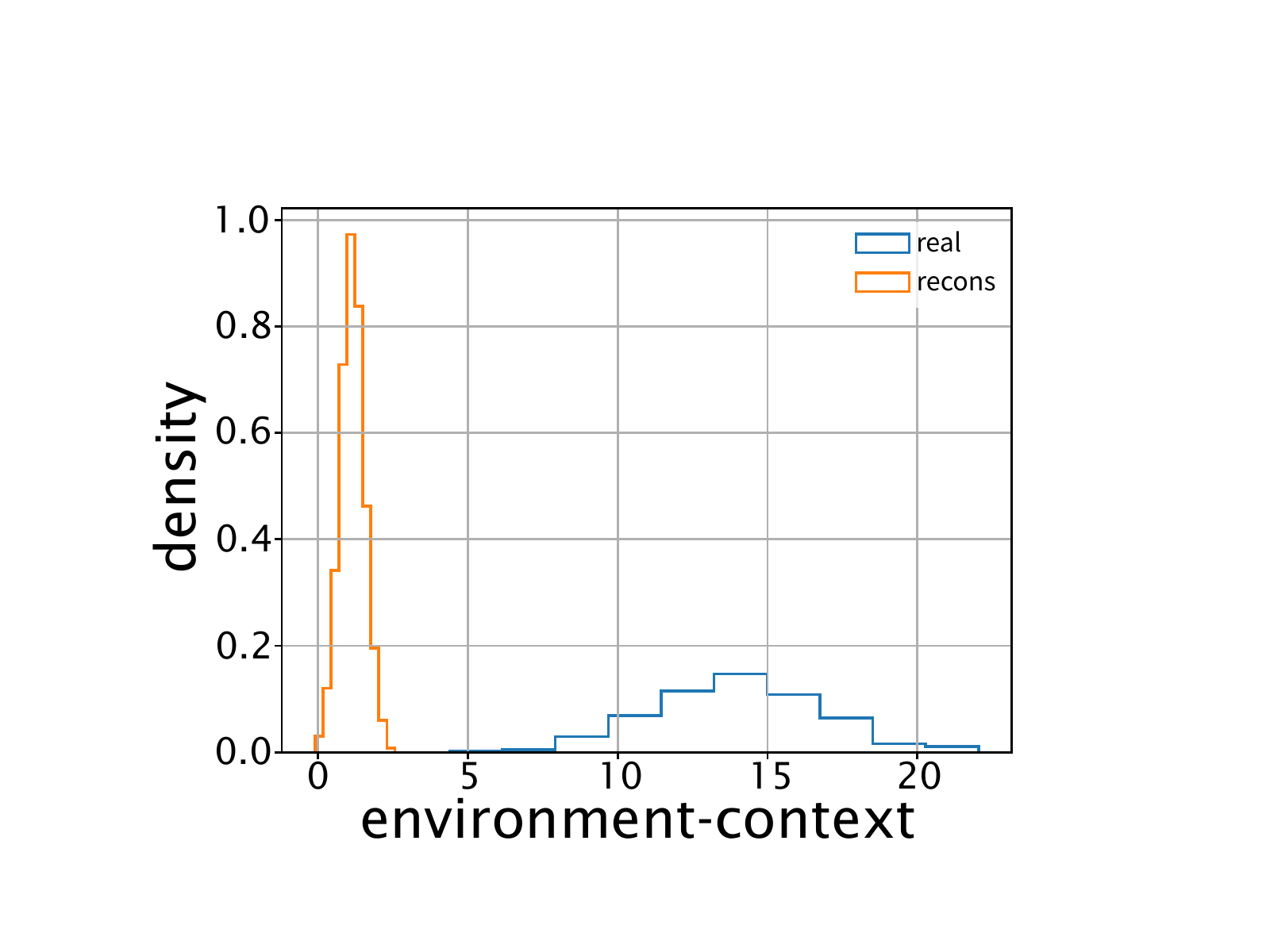}
	}

	\subfigure[histogram in the traning set ($\mu_c=6$) at epoch 8000]{
		\includegraphics[width=0.45\linewidth]{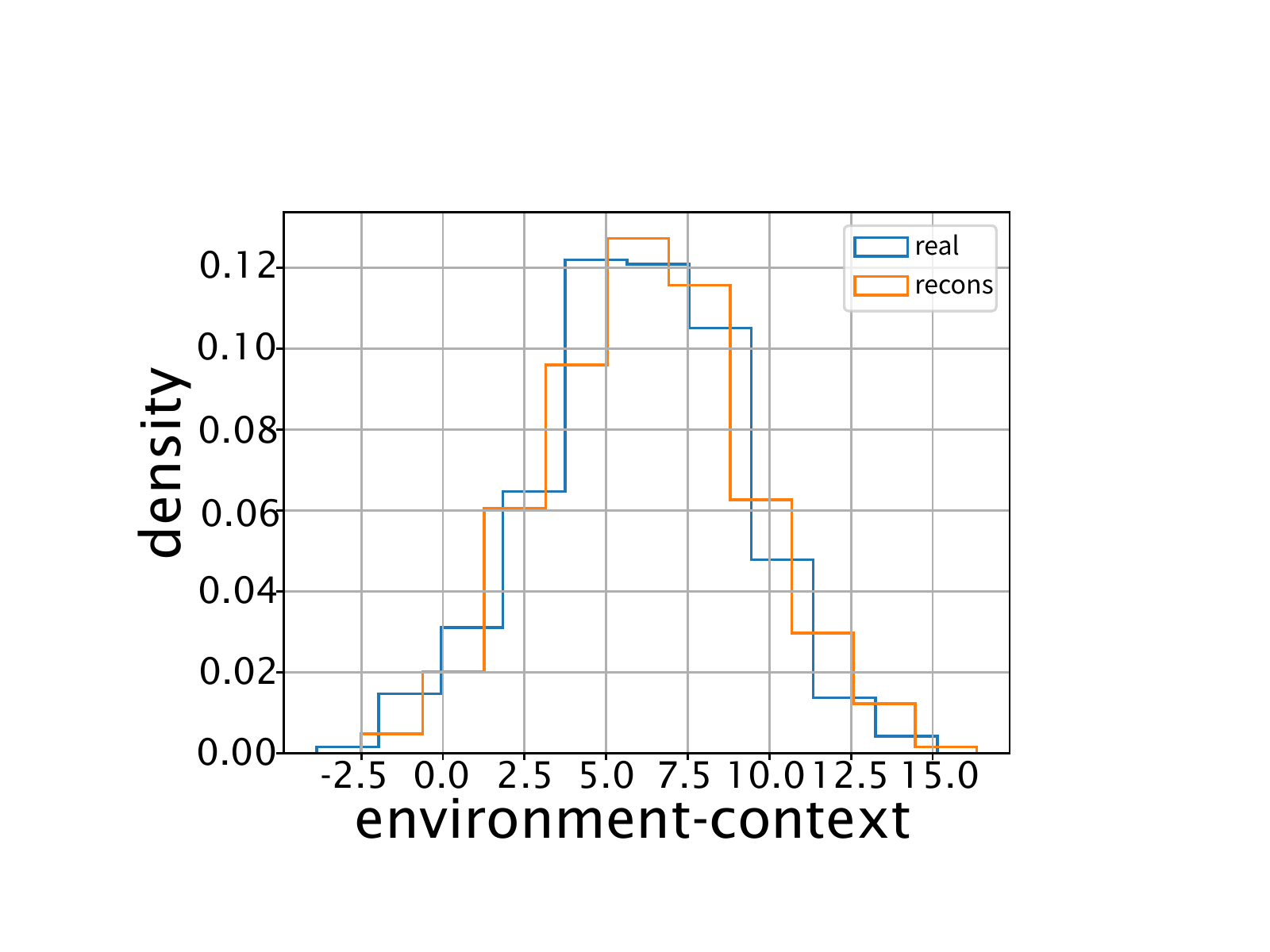}
	}
	\subfigure[histogram in the testing set  ($\mu_c=14$) at epoch 8000]{
		\includegraphics[width=0.45\linewidth]{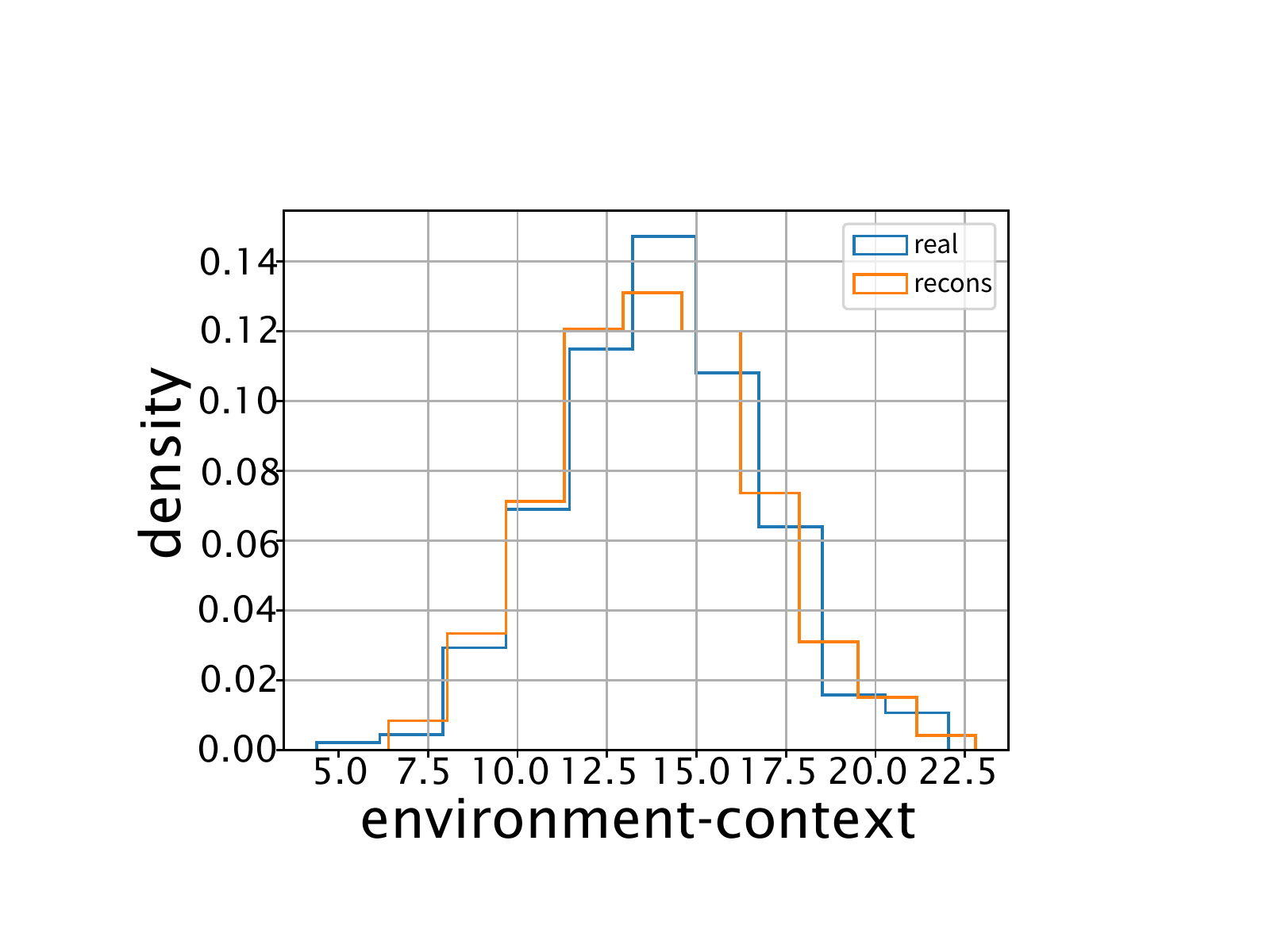}
	}
	\caption{Illustration of the histogram about user feature of $o^i$ in reconstructed and real data in the task of LTS3.}
	\label{fig:lts:vae-rec}
\end{figure}

We use KLD to measure the performance of reconstruction.  Since $p_\theta\left(\psi_s|\upsilon\right)$ also outputs the parameters of Gaussian distribution, we compute the KLD directly via the analytic expression of Gaussian distribution between $p_\theta\left(s|\upsilon\right)$ and $ \mathcal{N}(\mu_c, \omega_c)$.  We test the KLD every 100 epochs. Figure~\ref{fig:lts:vae-kld} shows that the KLD in the testing set finally converges to the range of 0.01 to 0.02. Figure~\ref{fig:lts:vae-rec} shows the reconstruction distribution is also correlated.

Finally, we analyze the embedding performance of SADAE by principal component analysis (PCA)~\cite{pca}. 
We first train $q_\kappa$ with a pre-collected dataset $\mathcal{D}$ and conduct PCA. The cumulative energy ratio of PCA in Fig.~\ref{fig:lts:pca} shows that: after 6000 epochs, the latent code can be almost represented by the first principal component. By projecting $\upsilon$ into two-dimensional vectors based on the first two principal components and comparing it with the ground-truth $\omega_g$, we can see that the value of $\omega_g$ linearly depends on the first principal component (See Appendix.~\ref{app:pca} for details).

	\begin{figure*}[h]
		\centering
		\vspace{-10mm}
		\subfigure[LTS1]{	\includegraphics[width=0.3\linewidth]{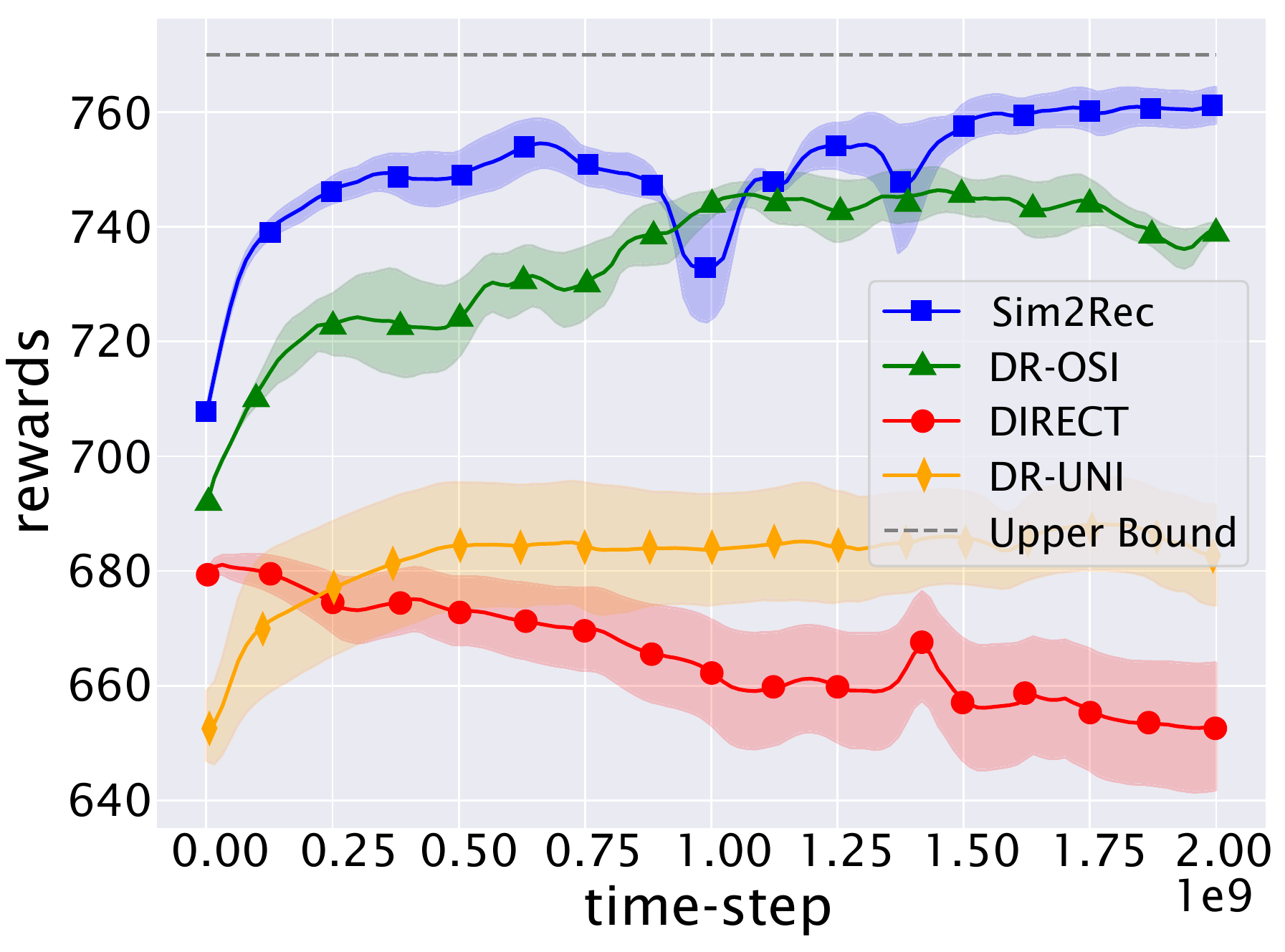}
			\label{fig:lts2}
		}
		\subfigure[LTS2]{
			\includegraphics[width=0.3\linewidth]{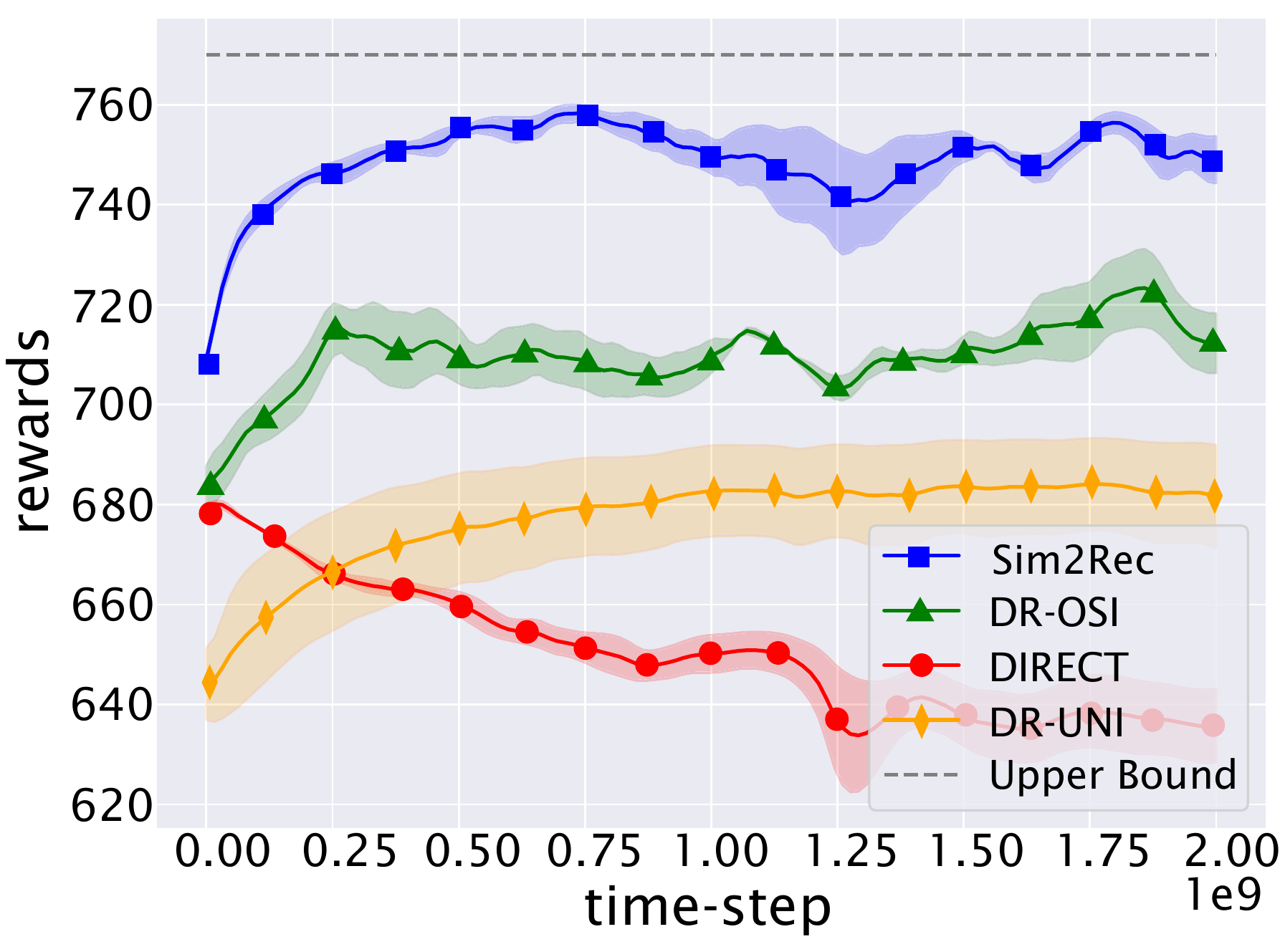}
			\label{fig:lts3}
		}
		\subfigure[LTS3]{	\includegraphics[width=0.3\linewidth]{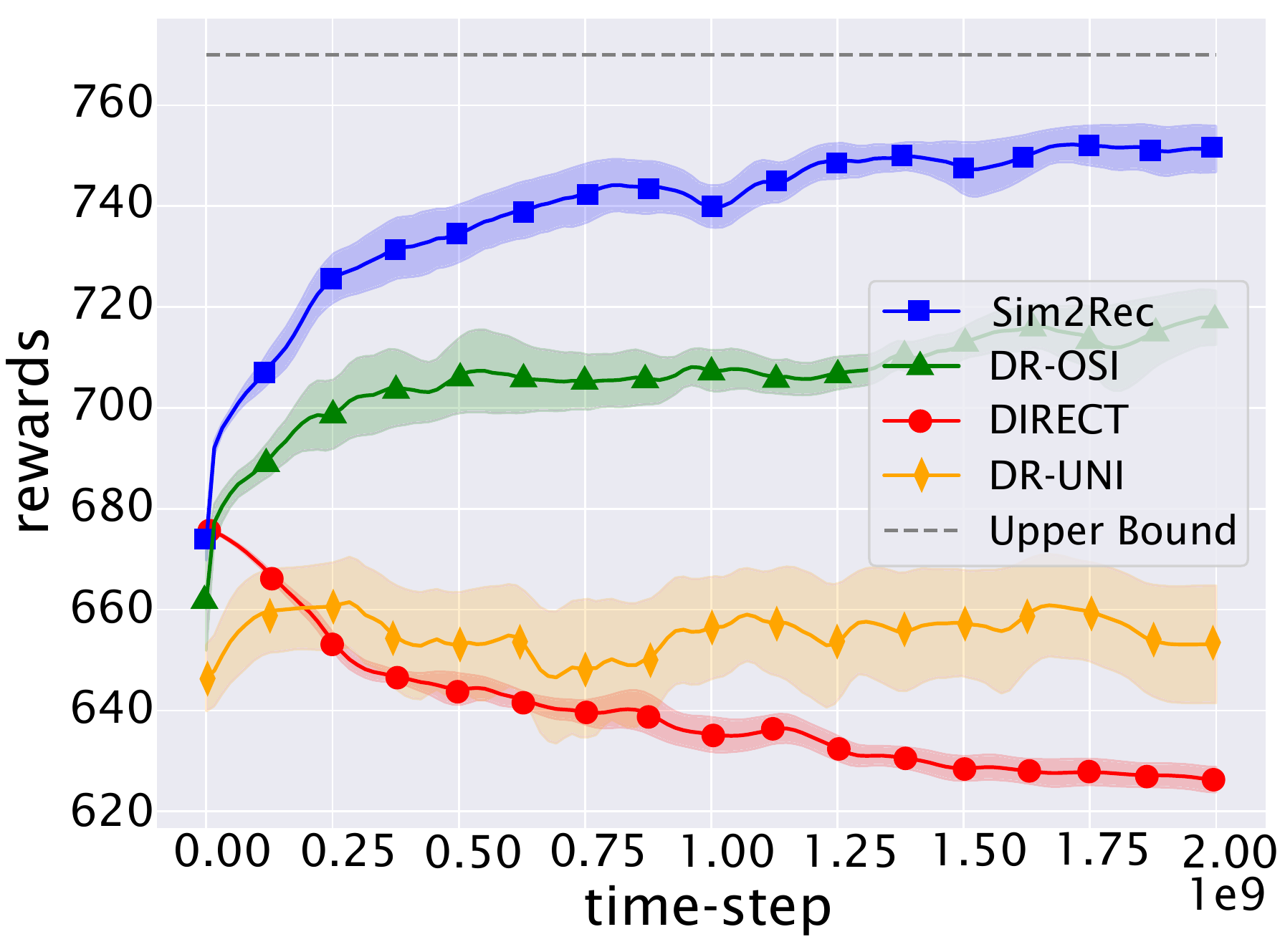}
			\label{fig:lts4}
		}
		\hspace{-10mm}
		\caption{Illustration of the performance in synthetic environments. The solid curves are the mean reward and the shadow is the standard error of three seeds. ``Upper Bound'' is the performance of a policy trained in the target domain directly. We regard it as the upper bound performance.}
		\label{fig:lts}
		\vspace{-5mm}
	\end{figure*}
\subsubsection{Results of the Policy Performance  (\textbf{RQ2})}
\label{exp:policy-syn}

We then test the adaptability of \our{} in SRS. 
We report our results in Fig.~\ref{fig:lts}.  
First, the results of DIRECT show that the performance degradation is severe in the tasks. Without considering the difference between training and deploying, the policy generates unpredictable behaviors. Second, all algorithms which consider learning from multiple dynamic models can improve the robustness in unknown environments. However, the algorithms that adopt  the representation of environments (\our{} and DR-OSI) reach better performance since they try to find the optimal policy in the representation of the environment instead of maximizing the expectation performance in the training set. In addition, \our{} reaches the near-optimal performance and does better than DR-OSI in difficult tasks (e.g., LTS3), which validates the necessity and effectiveness of SADAE proposed in Sec~\ref{sec:phi}. %
In more difficult tasks, the limitation of the representation ability bounds the performance of the context-aware policy.

\begin{figure}[h]
		\centering
		\subfigure[500-user simulators]{
			\includegraphics[width=0.46\linewidth]{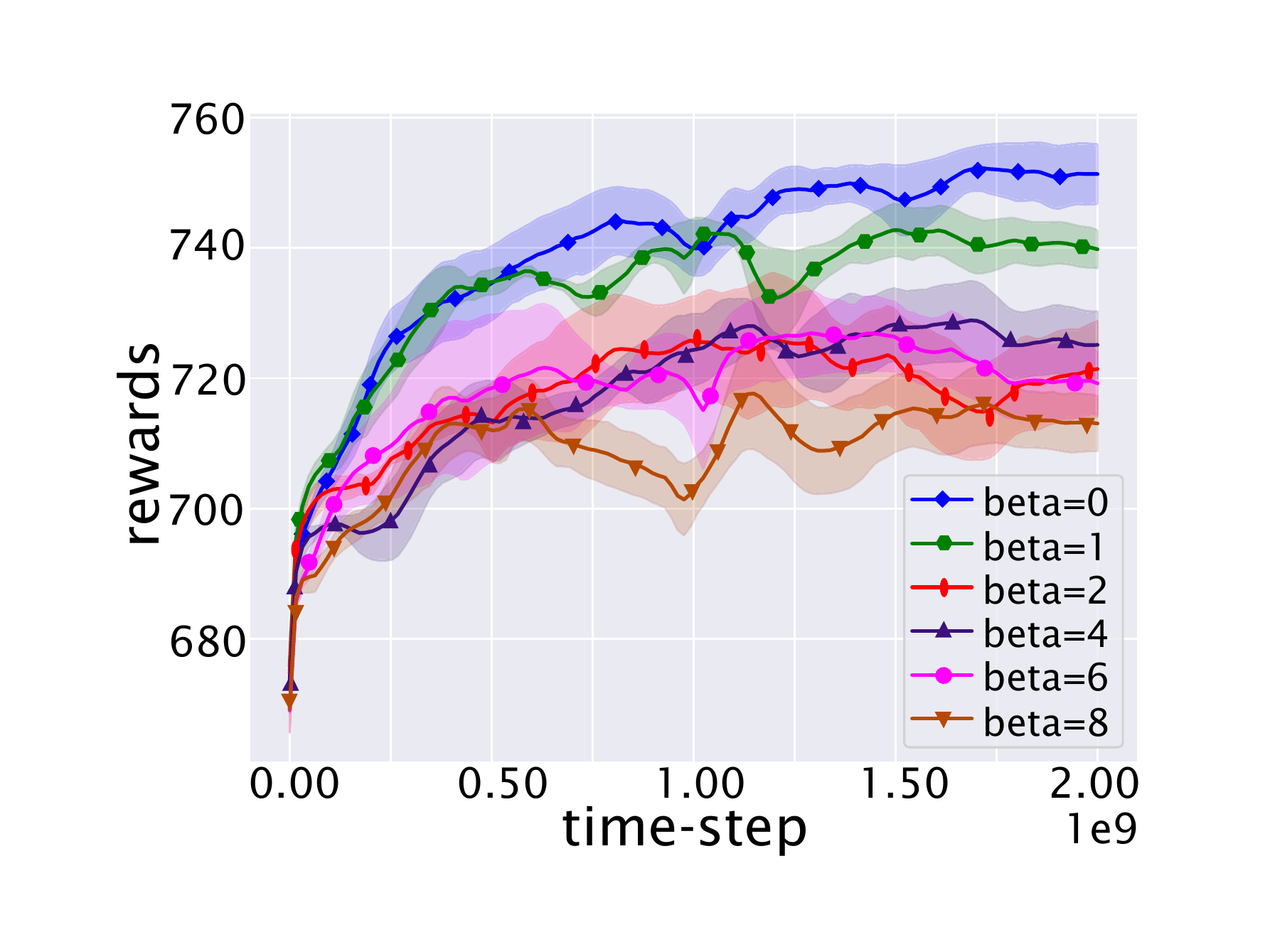}
			\label{fig:lts-limited}
		}
		\subfigure[unlimited-user simulators]{
			\includegraphics[width=0.45\linewidth]{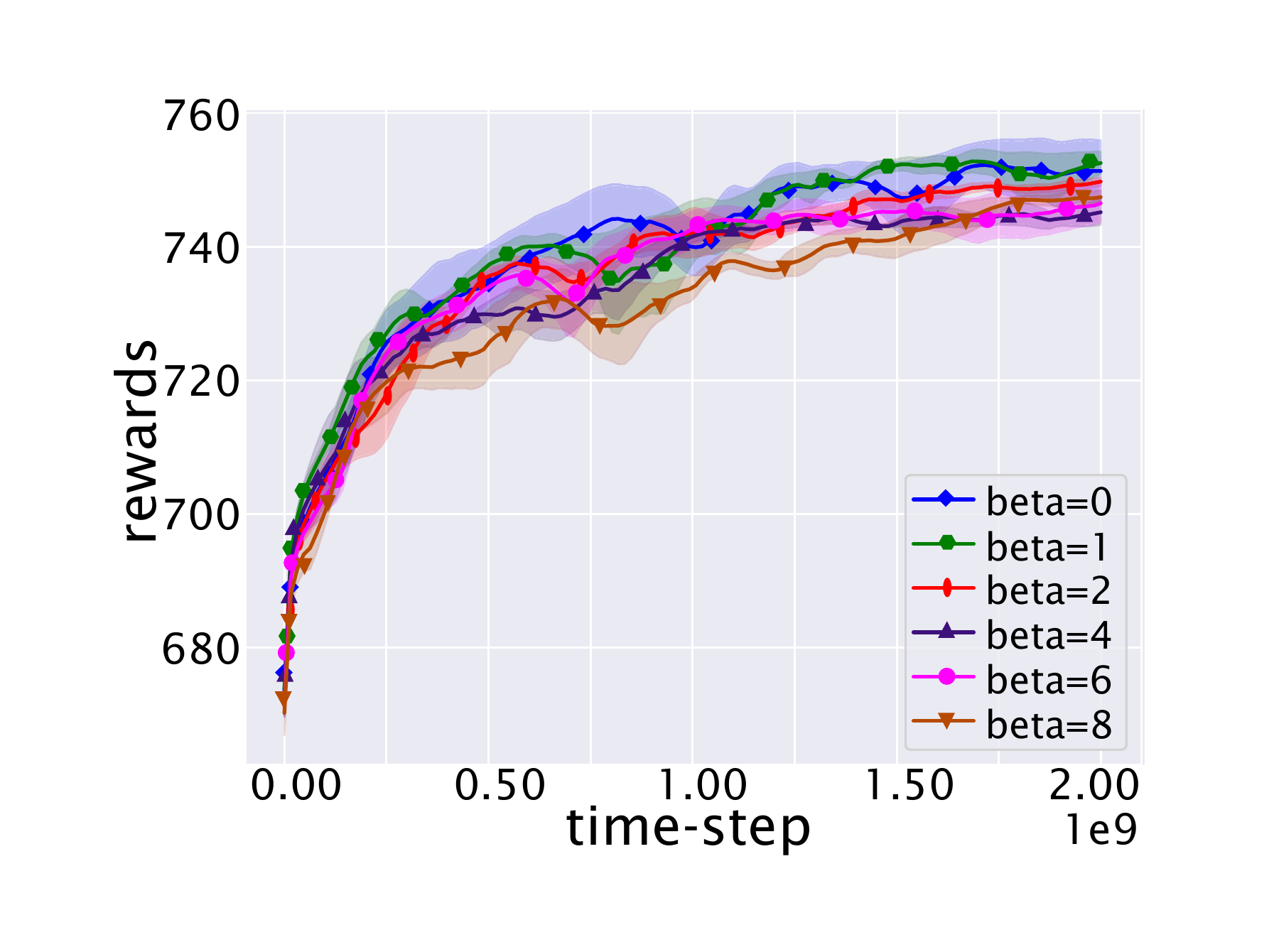}
			\label{fig:lts-unlimited}
		}
		\caption{Illustration of the performance in the LTS3-$\beta$ tasks. The solid curves are the mean reward and the shadow is the standard error of three seeds. In the 500-user simulators setting, we sample $\omega_\user$ from ${\rm Uni}(-\beta, \beta)$ for each simulator and each user. In the unlimited-user setting, we re-sample $\omega_\user$ for each simulator at each iteration of policy learning.}
		\label{fig:lts-limit-test}
	\end{figure}

	We finally analyze the influence of the coverage of simulator set on $\omega^*$. We conduct the experiment in LTS3-$\beta$, which inject parameter gaps for each user in the simulator. 
	Fig.~\ref{fig:lts-limit-test} shows the performance of \our{} in this setting. We can see in Fig.~\ref{fig:lts-limited} that the deployed performance of \our{} with limited training set declines when the gap level becomes larger, but the performance is still better than the compared methods. Besides, in Fig.~\ref{fig:lts-unlimited}, we find that with enough sampled simulators, \our{} can overcome the reality-gap problem well. In conclusion, empirically, with a simulator set that cover $\omega^*$, it is possible to overcome the reality-gap problem via \our{}.

\subsection{Experiments in a Real-World Application}
\subsubsection{Driver Program Recommendation (DPR) Tasks in DidiChuxing}
The goal of DidiChuxing is to balance the demand from passengers and the supply of drivers, i.e., helping drivers finish more orders, and satisfying the more trip demand from passengers.
Driver program recommendation (DPR) is a typical task of SRS in the ride-hailing platform. In DPR, to satisfy more demands from passengers, we would like to maximize the long-term engagement of drivers in different regions and cities via recommending reasonable item sequences from the programs. The engagement is characterized by the cumulative \textit{orders} completed by each driver. The selected programs are put to drivers once a day. The programs include two features: (1) tasks for the driver to accomplish, which is modeled by a continuous variable.  If a driver completes the recommendation program, his/her engagement would be increased in our platform; (2) The expenses of the platform when a driver completes a program, also as a bonus for driver;
As drivers respond differently to the same tasks in different regions, We should determine the best recommendation for the programs based on the preference of the drivers and the groups they belong to.

The DPR can be modeled as an MDP. For simplification, we assume the influence among drivers can be ignored. It is reasonable since drivers almost have no ideas about other drivers' tasks. In the DPR environment, we regard each day as a timestep. At timestep $t$,  the recommendation system policy $\pi$ sends a program $a^i_t=\pi(s^i_t)$ to driver $i$ based on the observed feature $s^i_t$. $a^i_t$ denotes the program features.

\subsubsection{Implementations} 

For SADAE, $q_\kappa\left(\upsilon|s^{(i)}, a^{(i)}\right)$ outputs the Gaussian distribution parameters of $\upsilon$. $p_\theta\left(\psi_a|\upsilon, s^{(i)}\right)$ and $p_\theta\left(\psi_s|\upsilon\right)$ output the parameters of the distributions. The action reconstruction is modeled with Gaussian distribution since it is continuous in the DPR. However, the state space includes continuous and discrete features. For simplification, we assume the continuous features are independent of discrete features. Thus we model them with Multivariate Gaussian distribution and categorical distribution respectively. The prior of $\upsilon$ is set to standard normal distribution, i.e., $p(\upsilon)= \mathcal{N}(0,1)$.

We reconstruct user simulators via DEMER~\citep{demer} which is a state-of-the-art user simulator learning techniques in ride-hailing platform.
As the simulator is built via a data-driven method, we adopt the proposed techniques in Sec.~\ref{sec:reliable} for feasible parameter space construction. The implement are as follows: (1) We train 15 simulators based on DEMER with different random seeds and different data sources of cities to construct $\Omega'$;
(2) For each time-step $t$, the reward penalty $U(s_t,a_t) = \mathbb{E}[ \Vert \mu_j(s_t,a_t) -\bar \mu(s_t,a_t) \Vert_2]$, where $\mu_j(s_t,a_t)$ denotes the mean of the predicted Gaussian distribution of the $i$-th simulator at state $s_t$ and action $a_t$, $\bar \mu(s_t,a_t)$ denotes the expectation of the simulators' predictions, and $\Vert \cdot\Vert _2$ denotes the l2-norm; (3) $T_c$ is set to 5 for all of our experiments in DPR; (4) $F_{\rm trend}$: we conduct an intervention test as the experiment in Fig.~\ref{fig:action_dist} and remove the drivers which the slope of reaction is negative or zero among all simulators; (5) $F_{\rm exec}$: we compute the minimal and maximal action values in the past 14 days for each driver in each group as the executable action subspace and adopt $F_{\rm exec}$ directly.

Finally for each time-step, the reward is set to:
$$
{\rm order} - {\rm cost} \times \alpha_1 - 0.01 \times U(s,a),
$$
where $\rm order$ is the finished order of the driver, $\rm cost$ is the expenses of the driver, which can be computed by $\rm order$ and $a$, $\alpha_1$ is a trade-off coefficient, which is set to the average GMV per order in the platform.

\subsubsection{The Offline Test Setups}

To conduct the offline test, we use 12 of the simulators in $\Omega'$ and 80\% data in the dataset for policy learning and the left simulator and data for testing. By selecting 3 of the simulators in $\Omega'$, named SimA, SimB, and SimC, as the deployment environment, we construct 3 tasks for testing. The tasks shared the same training and testing dataset.

\subsubsection{Group Information Reconstruction in Real Data  (\textbf{RQ1})}
\label{exp:sadae-real}
We train the SADAE in the training set and test the reconstructed data distribution in the unseen environment. The training dataset $\dataset$ comes from human expert data in the training set. 

\begin{figure}[h]
	\vspace{-5mm}
	\centering
	\subfigure[]{
		\includegraphics[width=0.45\linewidth]{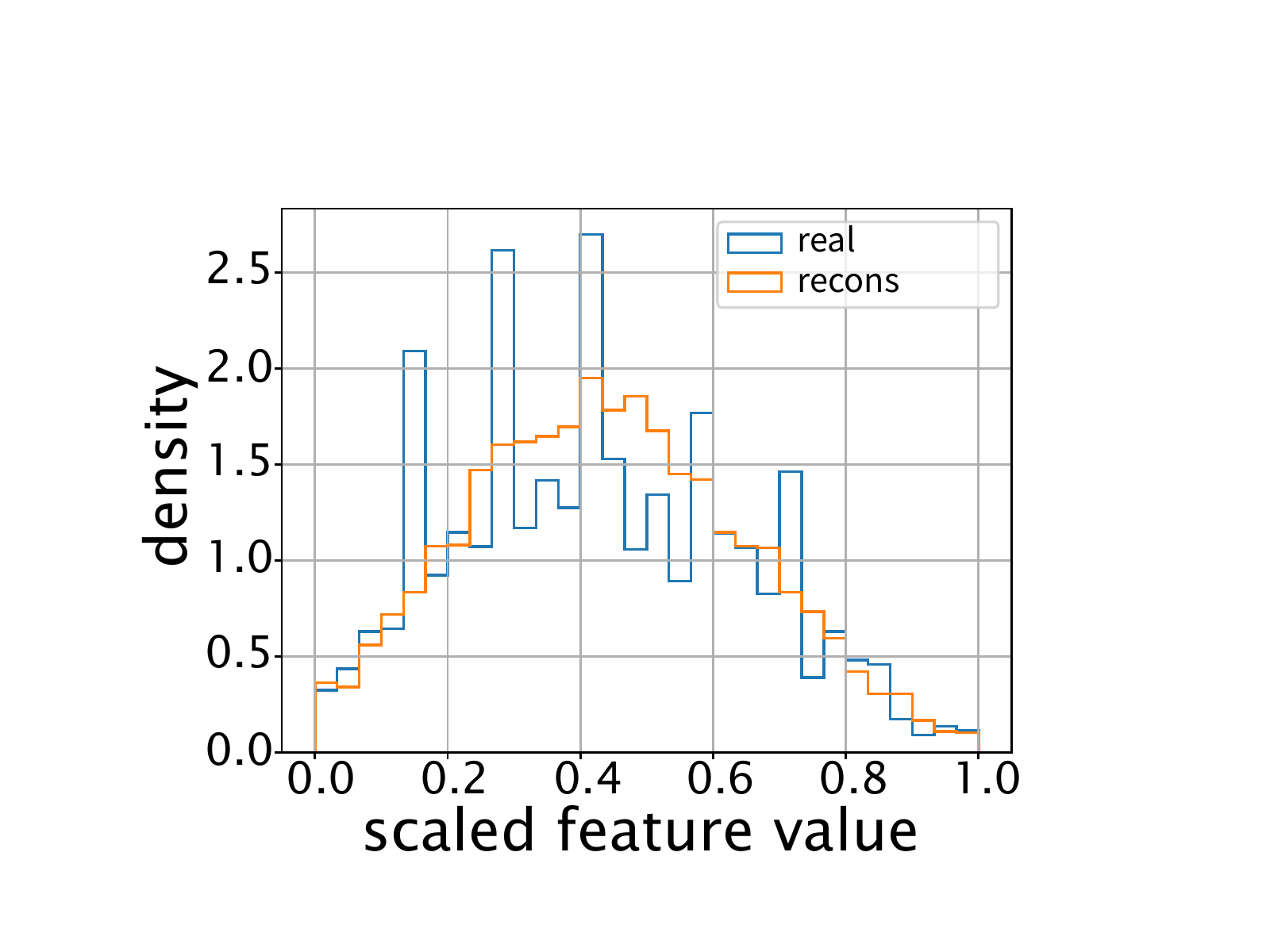}
	}
	\subfigure[]{
		\includegraphics[width=0.45\linewidth]{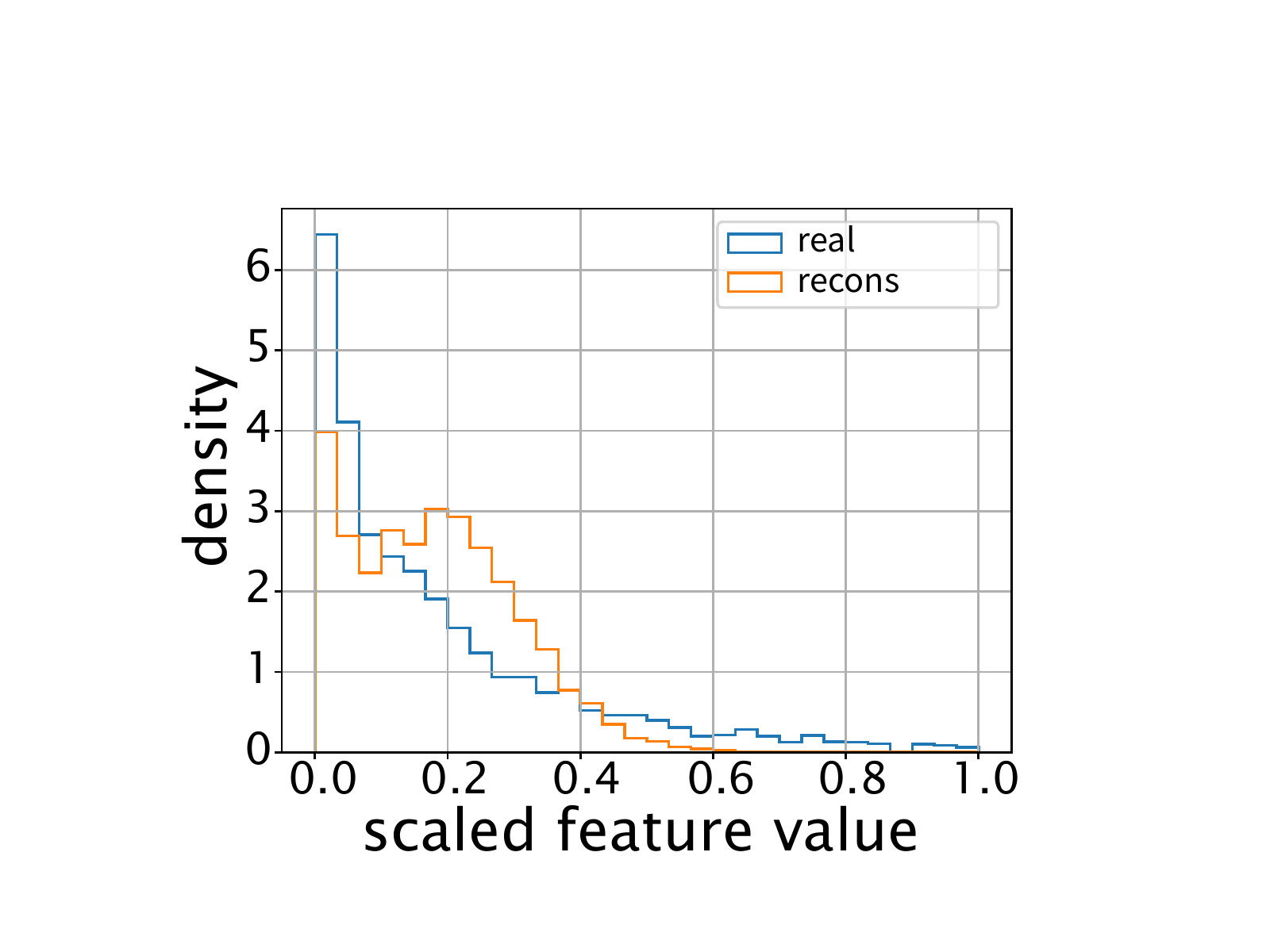}
	}

	\subfigure[]{
		\includegraphics[width=0.45\linewidth]{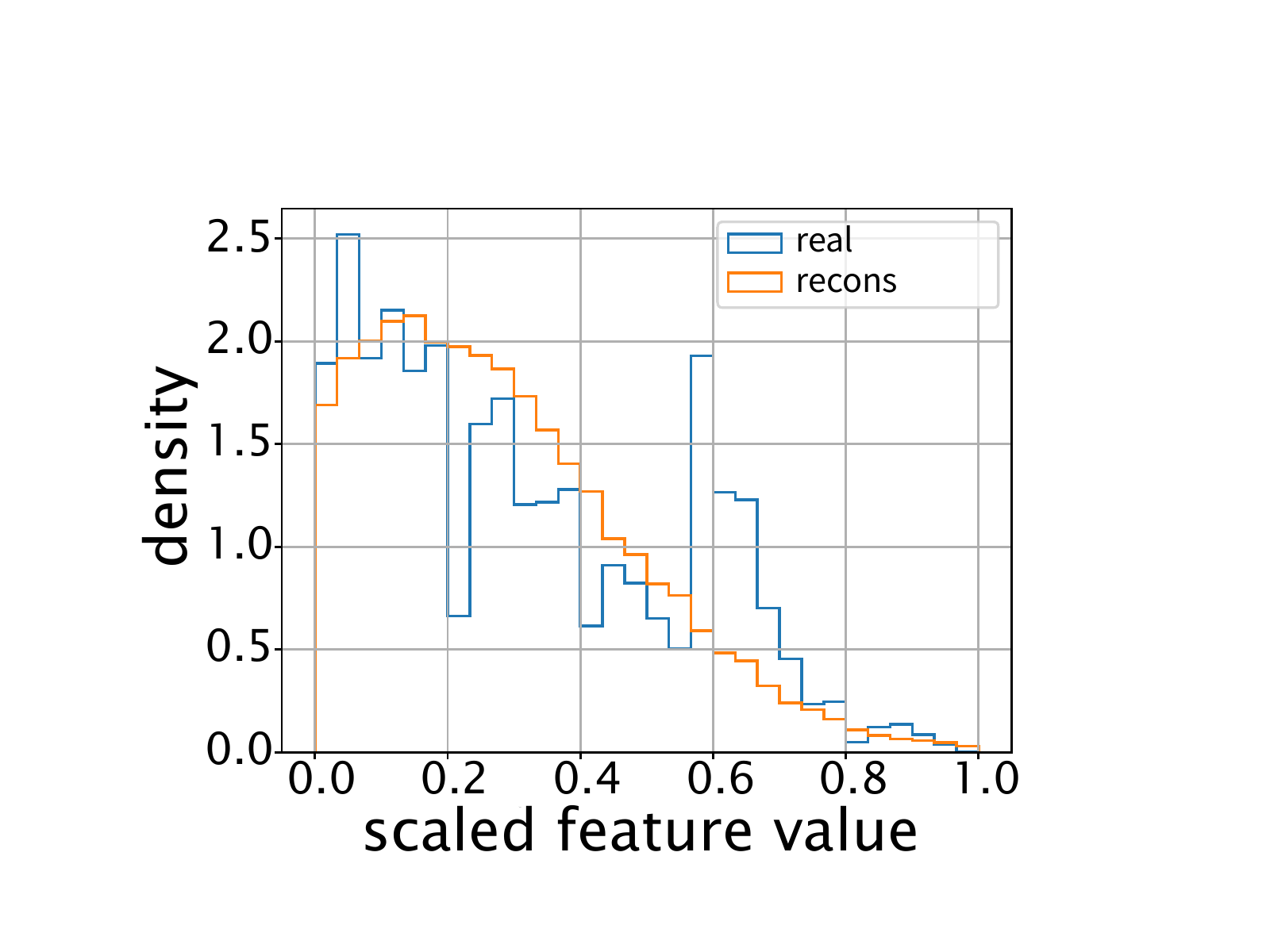}
	}
	\subfigure[]{
		\includegraphics[width=0.45\linewidth]{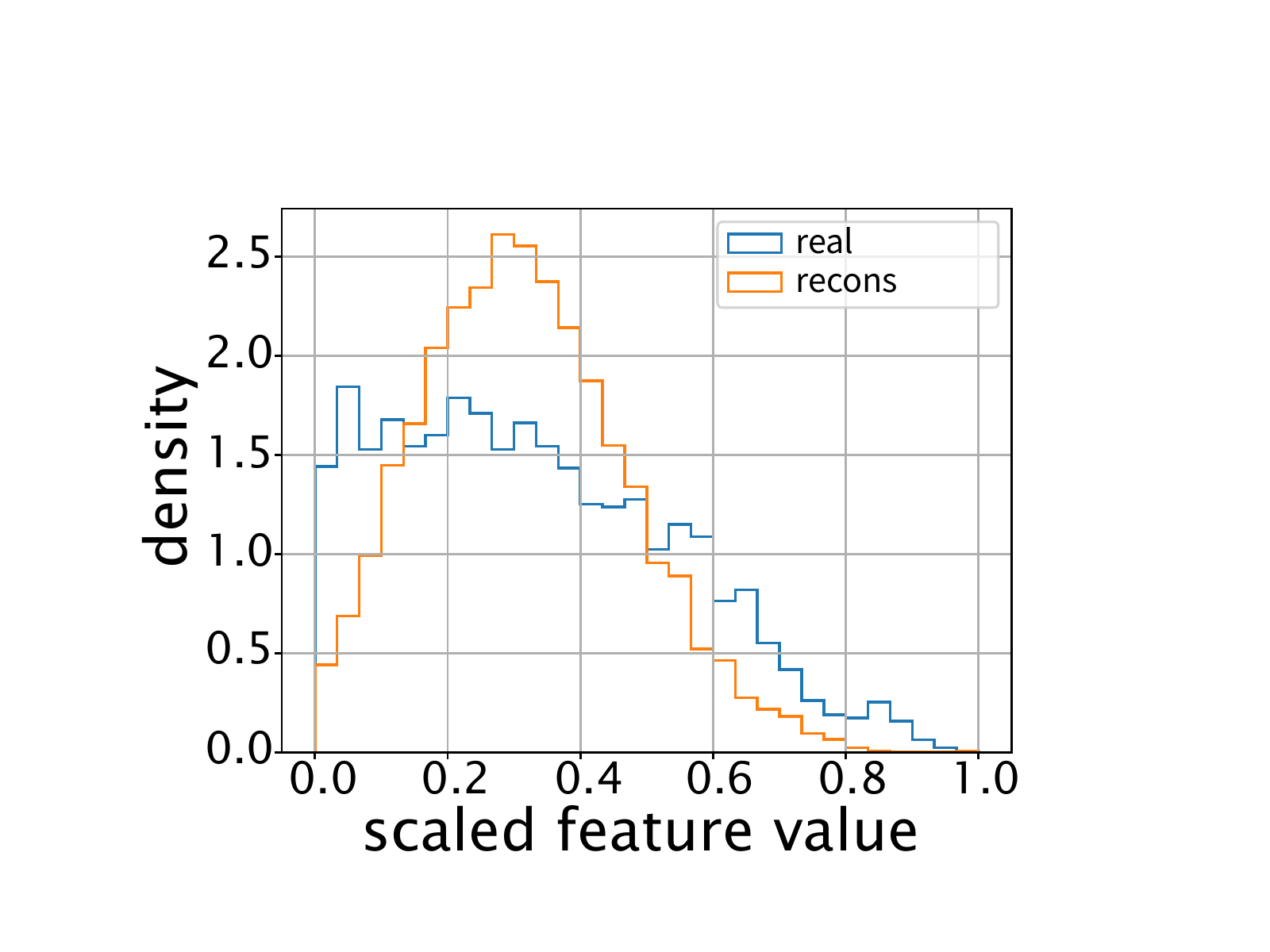}
	}
	
	\subfigure[]{
		\includegraphics[width=0.45\linewidth]{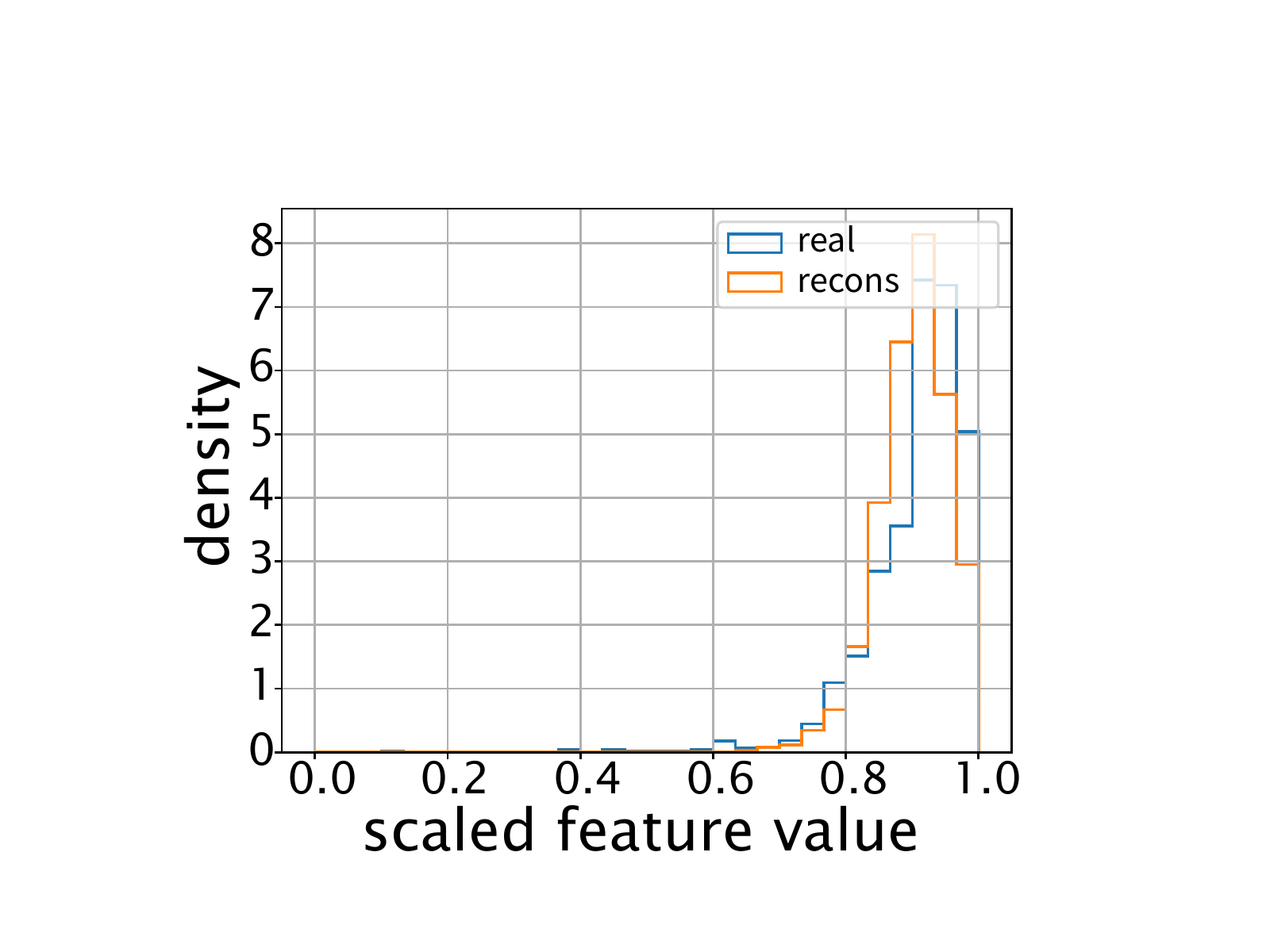}
	}
	\subfigure[]{
		\includegraphics[width=0.45\linewidth]{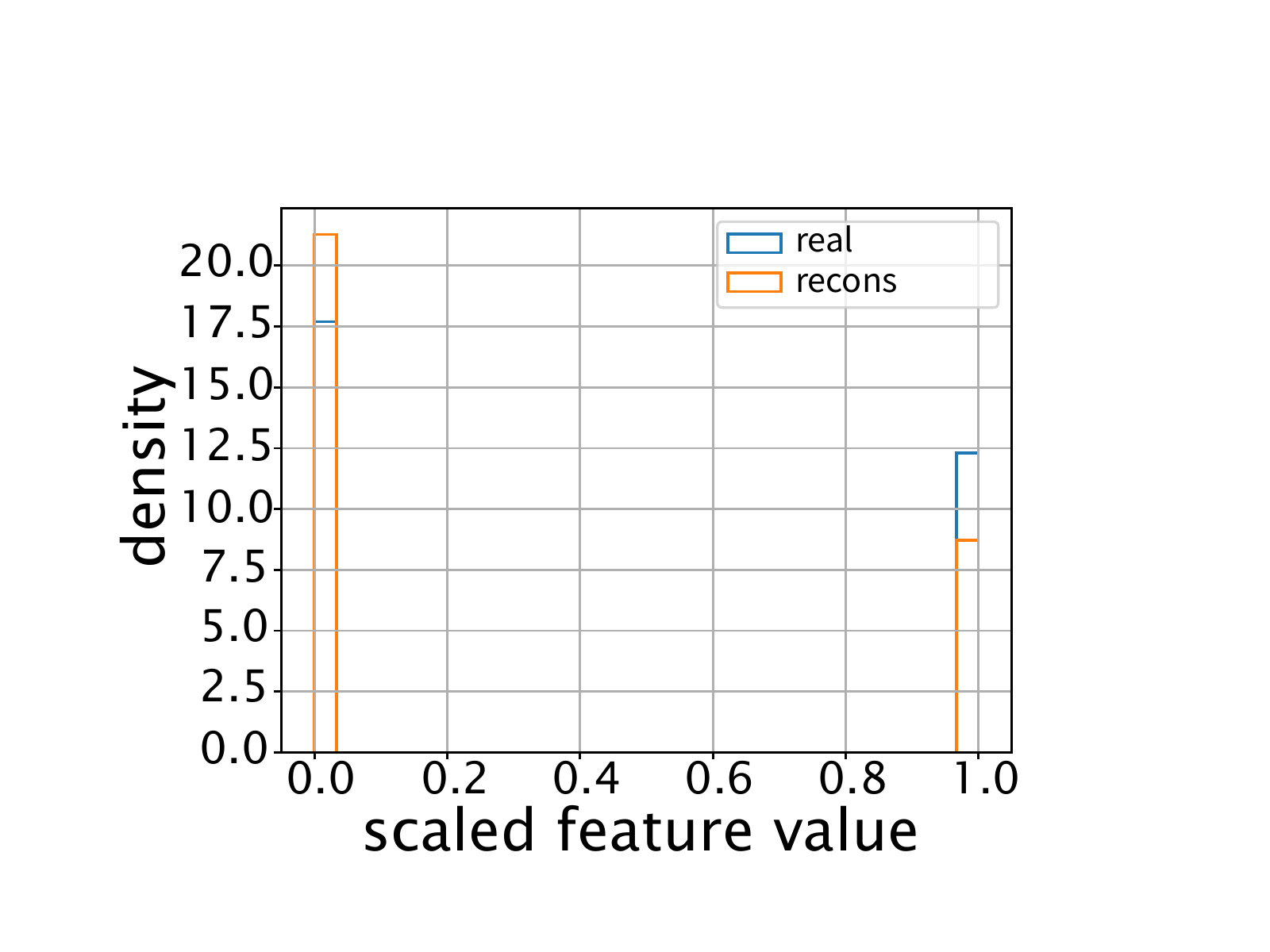}
	}

	\caption{Illustration of the histogram of reconstructed and real data on parts of the states.}
	\label{fig:rec}
\end{figure}

\begin{figure}[h]
	\vspace{-3mm}
	\centering
	\subfigure[KL divergence]{
		\includegraphics[width=0.45\linewidth]{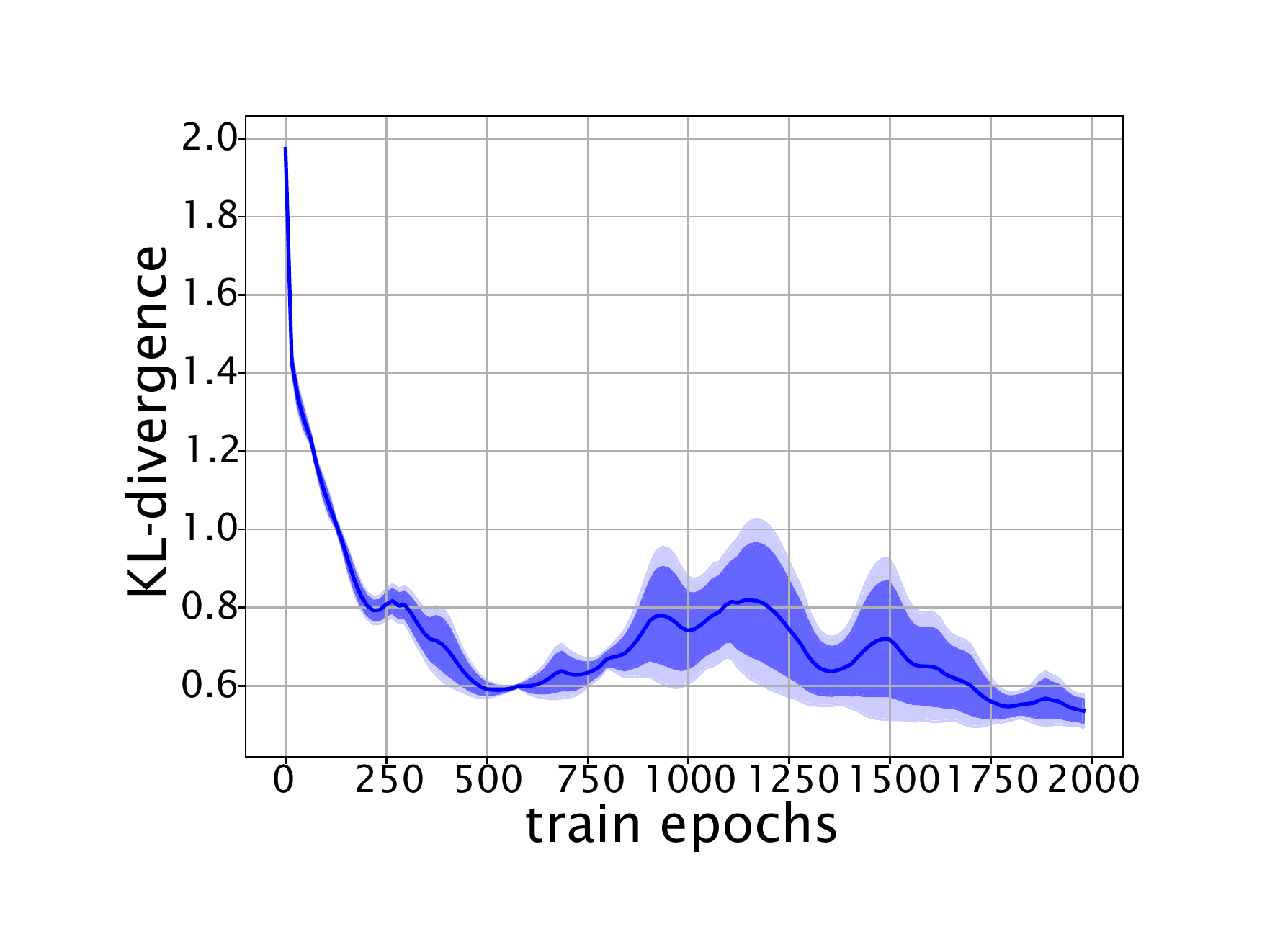} 
		\label{fig:kld}
	}
	\subfigure[mean absolute error]{
		\includegraphics[width=0.45\linewidth]{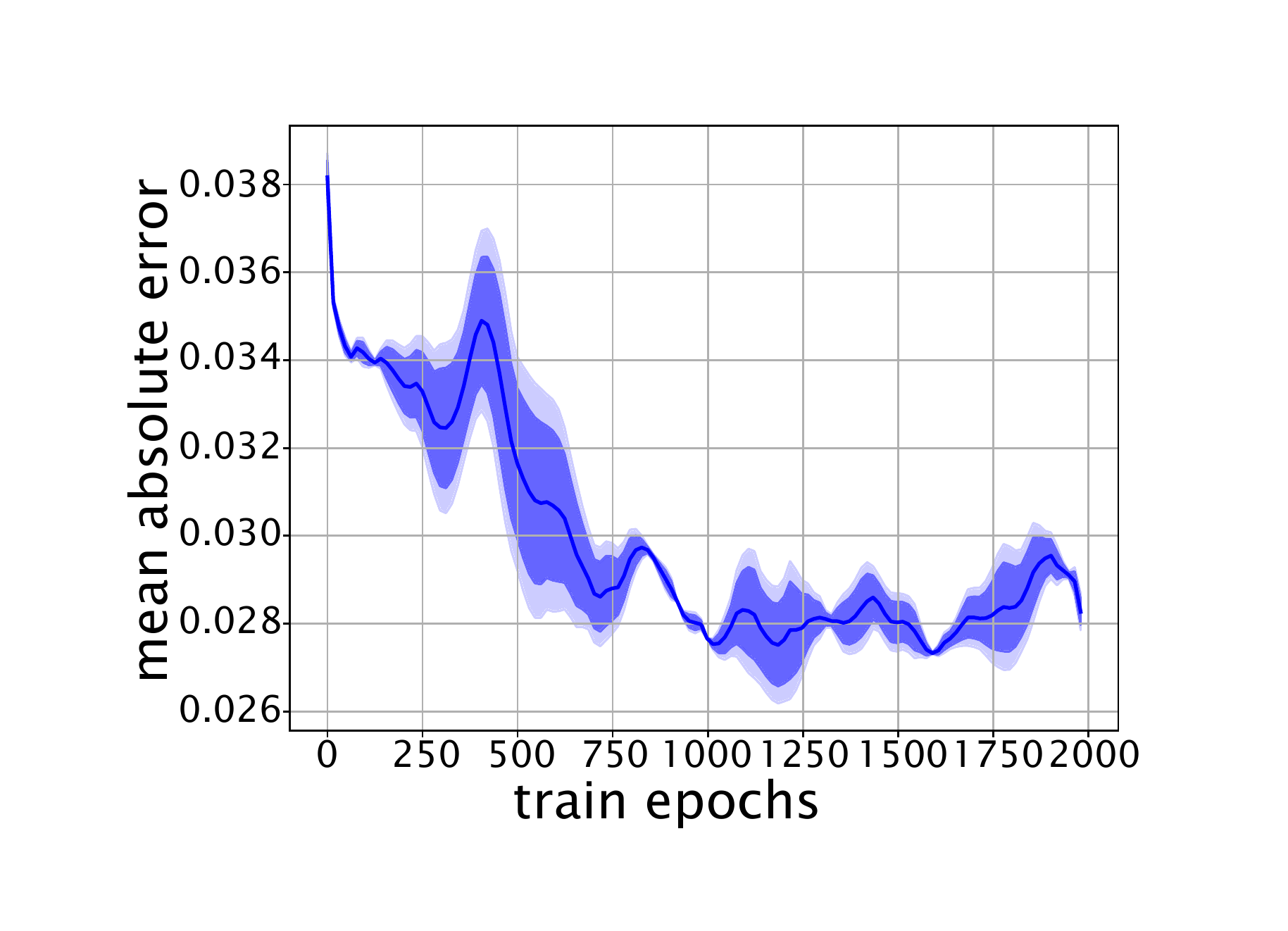} 
		\label{fig:mae}
	}
	\caption{Illustration of the reconstruction (left) and embedding (right) performance on SADAE. The solid curves are the mean value. The dark shadow is the standard error, while the light one is the min-max range of three seeds. } %
	\vspace{-5mm}
\end{figure}

We test the KLD every 100 epochs. Figure~\ref{fig:kld} shows that the KLD between the real data $X$ and the reconstructed distribution $p_\theta(X|\upsilon)$ steadily converges to 0.6, which demonstrates nontrivial reconstruction performance. Figure~\ref{fig:rec} shows histograms for examples of real and reconstructed data on a single feature, which are also significantly correlated. 

To evaluate the embedding performance of SADAE, we performed the hidden state prediction experiments~\cite{OpenAIcube}. We use another one-layer neural network to predict the KLD of two data pairs $(X_i, X_j)$, given their embedding variable $(\upsilon_i, \upsilon_j)$. The neural network has one 32-unit hidden layer with tanh as the activation function and links to a linear layer to predict the KLD computed by Eq.~\ref{equ:kld-compute}. The neural network is initialized and retrained for the same epochs, every 100 iterations of SADAE learning. If the embedding variables store useful information about the distribution, the KLD prediction error between arbitrary two datasets would be negatively correlated with the training epochs.
Fig.~\ref{fig:mae} shows the mean absolute error (MAE). The MAE has 26\% improvement than the initial variable, which implies the embedding variable is helpful to infer the relation of two distributions. 

\subsubsection{Necessity for the Feasible Parameter Space Construction  (\textbf{RQ3})}
\label{exp:param-space}

\begin{table}[ht]
	\caption{The performance of policies  learned with different policy learning techniques. The performance is tested in SimA.}
	\centering
	\begin{tabular}{ccccc}
		\toprule
		 & orders (test) & orders (train)  & cost (test) & cost (train)\\
		\midrule
		\our{} & 2.0\% &  1.6\% & 0.9\%  & 4.5\%  \\
		\our{}-PE & 1.3\% &  2.3\% & -8.0\% & -4.0\% \\
		\our{}-EE & 8.1\% &  8.2\% & -10.0\%  & -11.1\% \\
		\bottomrule
	\end{tabular}
    \label{tab:reality-gap}
\end{table}

\begin{figure}[ht]
	\centering
	\hspace{-10mm}
	\subfigure[SimA]{	\includegraphics[width=0.32\linewidth,height=0.285\linewidth]{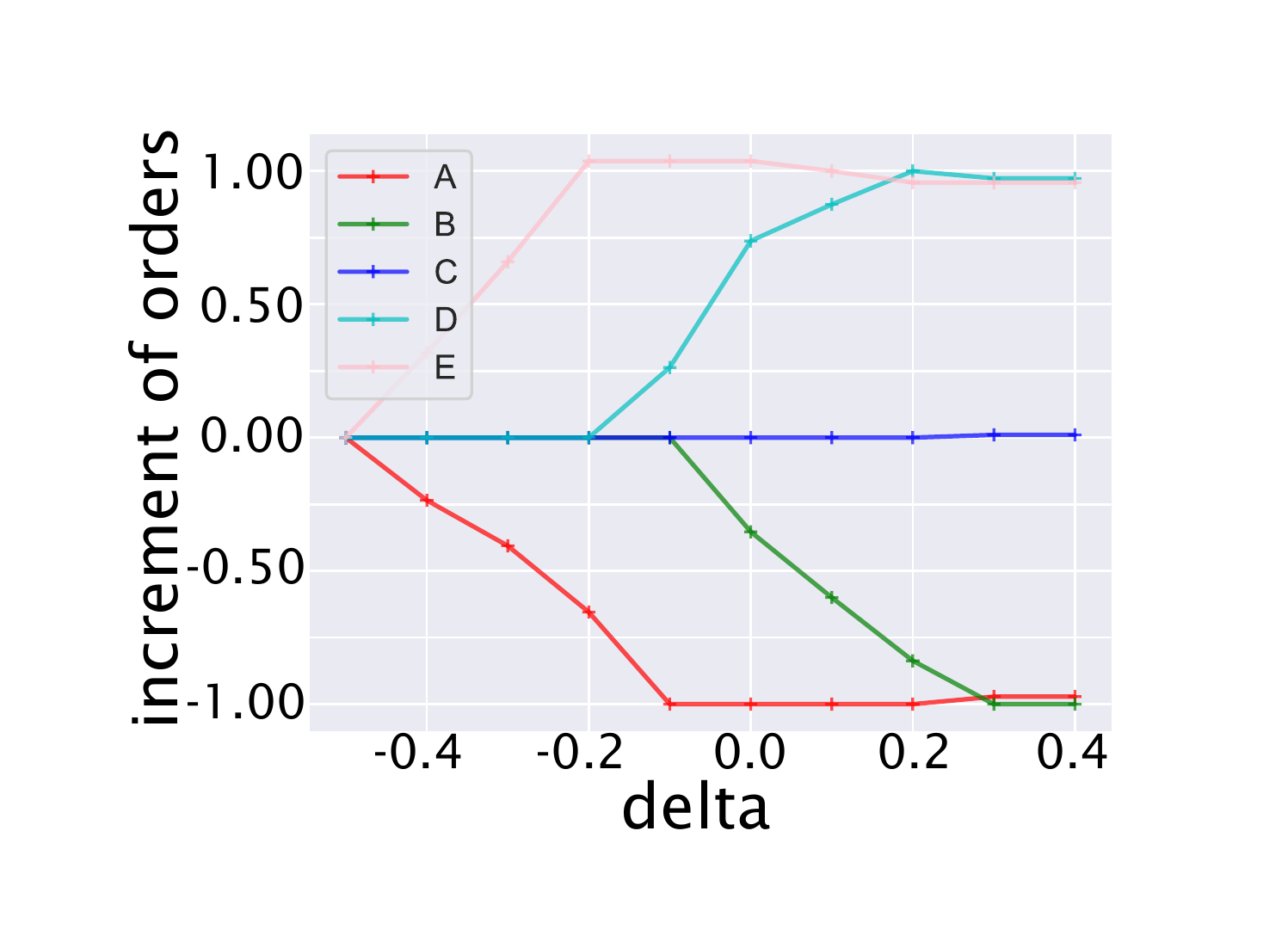}
		\label{fig:simA-reality-gap}
	}
	\subfigure[SimB]{
		\includegraphics[width=0.32\linewidth,height=0.285\linewidth]{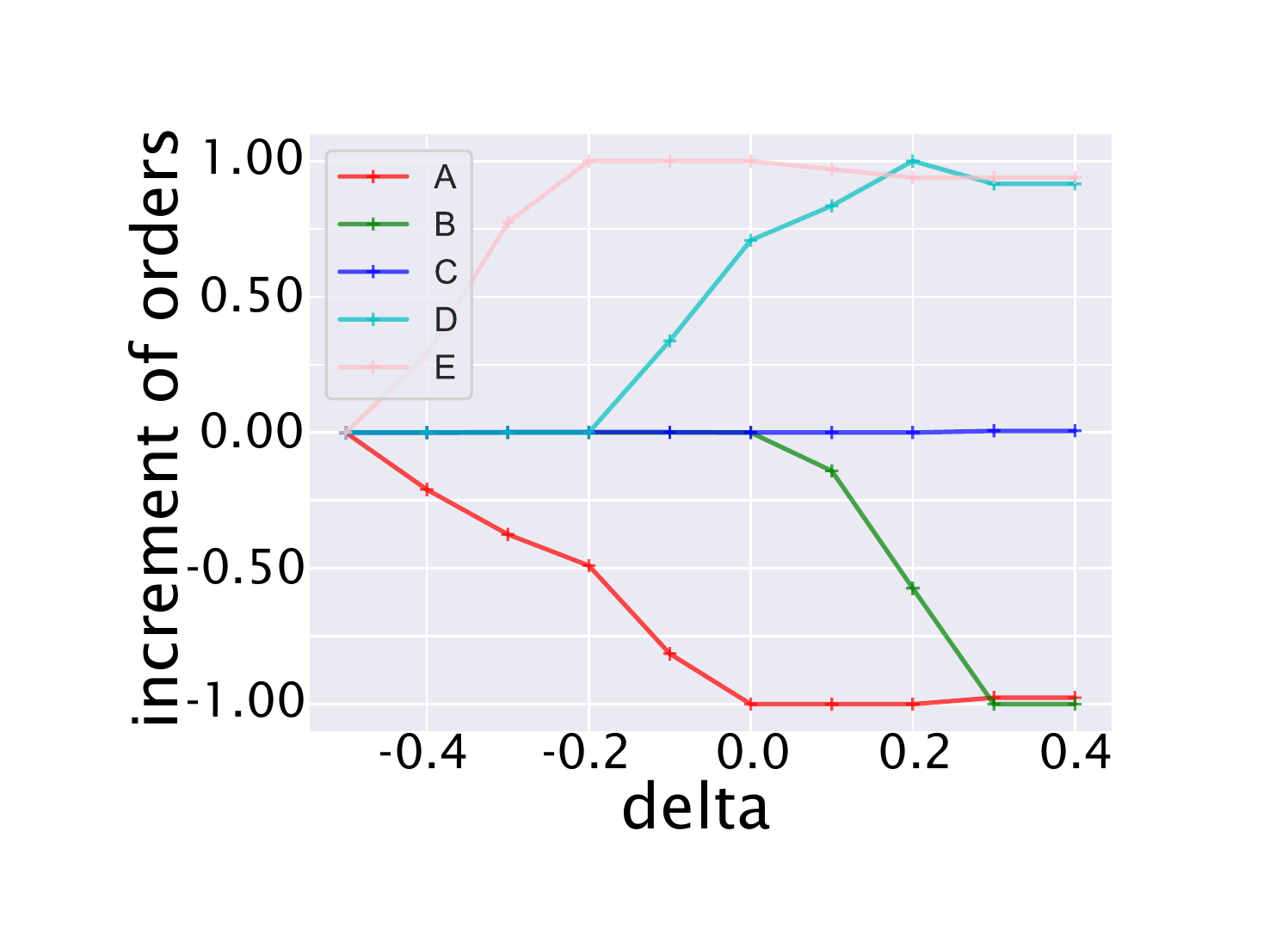}
		\label{fig:simB-reality-gap}
	}
	\subfigure[SimC]{	\includegraphics[width=0.32\linewidth,height=0.285\linewidth]{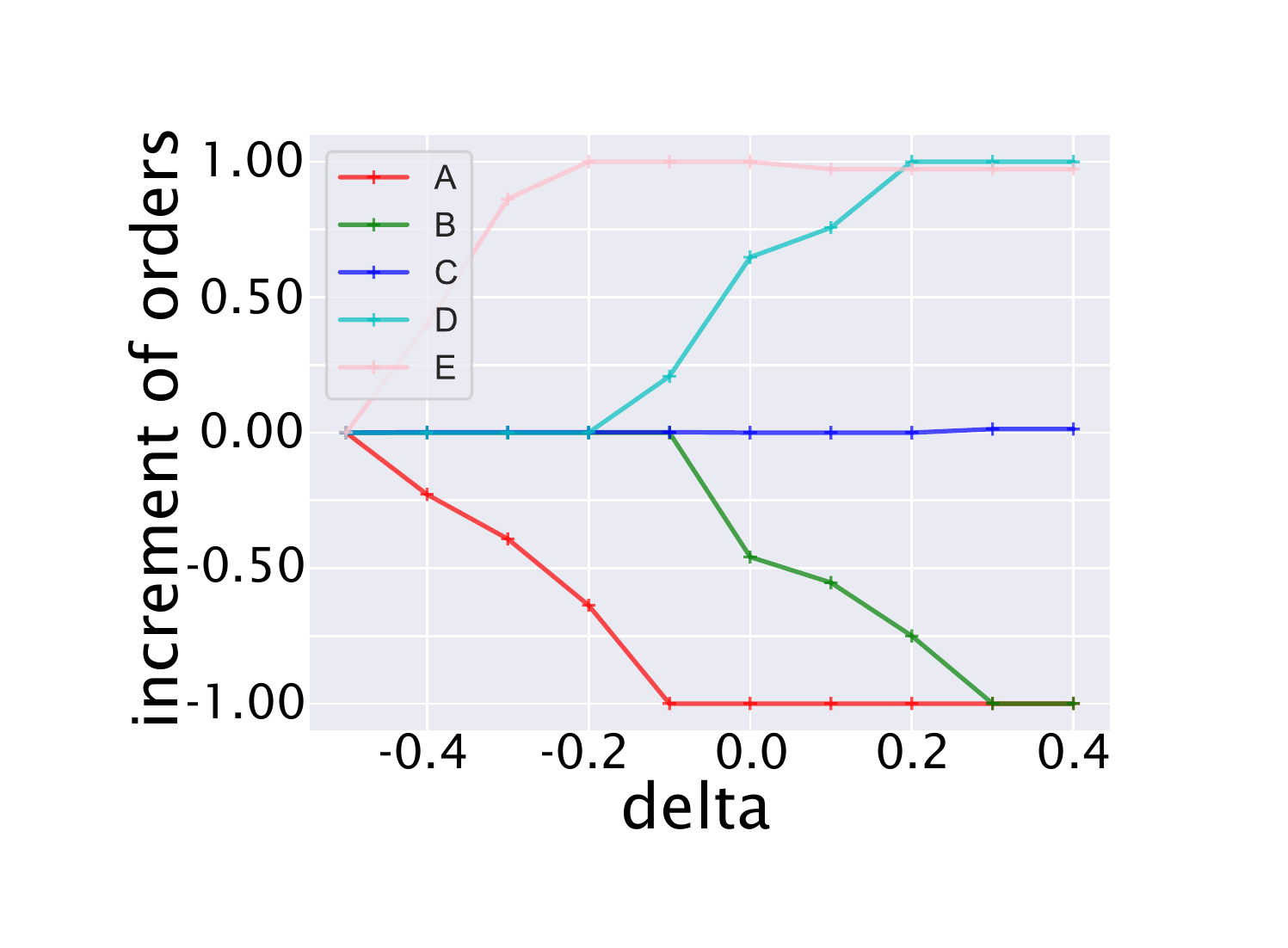}
		\label{fig:simC-reality-gap}
	}
	\hspace{-10mm}
	\caption{Illustration of the increment of orders on intervention test. Each figure plots the clustering centers of the drivers' response vectors in a simulator. Each line denotes a cluster center. The X-axis is the value of $\Delta B$. The increment of orders of each point is subtracted to the value in $\Delta B=-0.5$ of the corresponding cluster.}
	\label{fig:action_dist}
	\vspace{-5mm}
\end{figure}

We first demonstrate the reality-gap problem of SRS based on the application and show the effect of the reality-gaps for policy training if we do not implement the techniques in Sec.~\ref{sec:reliable} for training.  We compare \our{}  with \our{}-PE and \our{}-EE and list the percentage of increment of orders and costs in the training and testing set compared with the behavior policy $\pi_e$ in the logged dataset in Tab.~\ref{tab:reality-gap}.
As can be seen in Tab.~\ref{tab:reality-gap}, in the training set, \our{} reaches the lower order increment than \our{}-PE and \our{}-EE. 
However, in the \our{}-PE setting, when deploying the policy trained in \our{}-PE setting, the policy faces large performance degeneration (43\%), while the performance of \our{} keep similar between training and testing. The phenomenon indicates that performance improvement of \our{}-PE comes from the exploitation of the prediction error of the simulators which cannot generalize to the testing environment, which also validates the necessity of the technique proposed in Sec~\ref{sec:reliable} for avoiding the policy exploiting the regions with large prediction errors.
On the other hand, policy trained in \our{}-EE setting reaches better performance than \our{} both in training and testing set and with significant lower costs. 
However, the improvement comes from the policy exploiting the extrapolation error of the simulators, which are common among these ensemble simulators.  
To demonstrate this, we conduct an intervention test on the simulators (Fig~\ref{fig:action_dist}). In the intervention test, we take the bonus $B$, which is one of the action, of each driver in the dataset as the original points and assign the bonus with the same bias $\Delta B$: $\bar B \leftarrow B +\Delta B$, then we record the prediction of feedback $Y$ of drivers based on the original state features and the bias bonus $\bar B$. For each driver, we concatenate $Y$ with different $\Delta B$  and group the response vectors into 5 different clusters via K-means, which is in Fig~\ref{fig:action_dist}. In the intervention test, we find that the reaction patterns are similar among different simulators and there are some patterns that violate the prior knowledge (e.g., A, B, and C). There are many drivers that will be in the same patterns among the simulators, for example, according to our statistics, there are $15\%$ of drivers always in cluster C among the simulators. The reaction violates the fact and will mislead the policy training to get an unreasonable high performance. The policy can reduce the bonus to get more engagement for the drivers in pattern A, which explains why \our{}-EE receives much larger orders and smaller costs. This also demonstrates the necessity of the method proposed in Sec~\ref{sec:reliable} to guarantee the policy optimizing in the regions without large extrapolation errors, so that the policy can be less mislead.

\subsubsection{Policy Performance in Offline Tests  (\textbf{RQ4})}
\label{exp:policy-real}

\begin{table}[h]
 \vspace{-2mm}
	\caption{The performance of policies learned with different algorithms. Here we use the expectation cumulative rewards among drivers as the metric of performance. }
	\label{tab:perf-demer}
	\centering
	\begin{tabular}{ccccc}
		\toprule
		 & SimA & SimB & SimC \\
		\midrule
		\our{} & \textbf{0.470} &  \textbf{0.483 }& \textbf{0.479} \\
		DIRECT & 0.450 & 0.241 & 0.027 \\
		DeepFM & 0.325  &  0.302 & 0.368 \\
		WideDeep & 0.192 &  0.398  & 0.211 \\
		\bottomrule
	\end{tabular}
    \label{tab:offline-test}
\end{table}

In the above discussion, we verified the necessity of the proposed techniques for policy learning. We now demonstrate the performance of \our{} based on the simulators. We compare \our{} with two recommender systems based on supervised learning methods: DeepFM~\citep{deepfm} and WideDeep~\citep{widedeep}, and DIRECT~\citep{demer}. The results are listed in Tab~\ref{tab:offline-test}. 
We find that the transfer performance decline in DeepFM is not significant. DeepFM and WideDeep can also get rewards from the logged dataset to some degree. We surmise that the RL-style algorithms, e.g., DIRECT, is more likely to overfit the simulator, leading to unreliable behavior when deployed~\cite{overfit}. However, in three tasks, \our{} always gets the optimal performance.

\subsection{AB Test in the Production Environment (\textbf{RQ5})}
\label{exp:ab}

\begin{figure}[ht]
    	\centering
    \vspace{-2mm}
    	\includegraphics[width=0.55\linewidth]{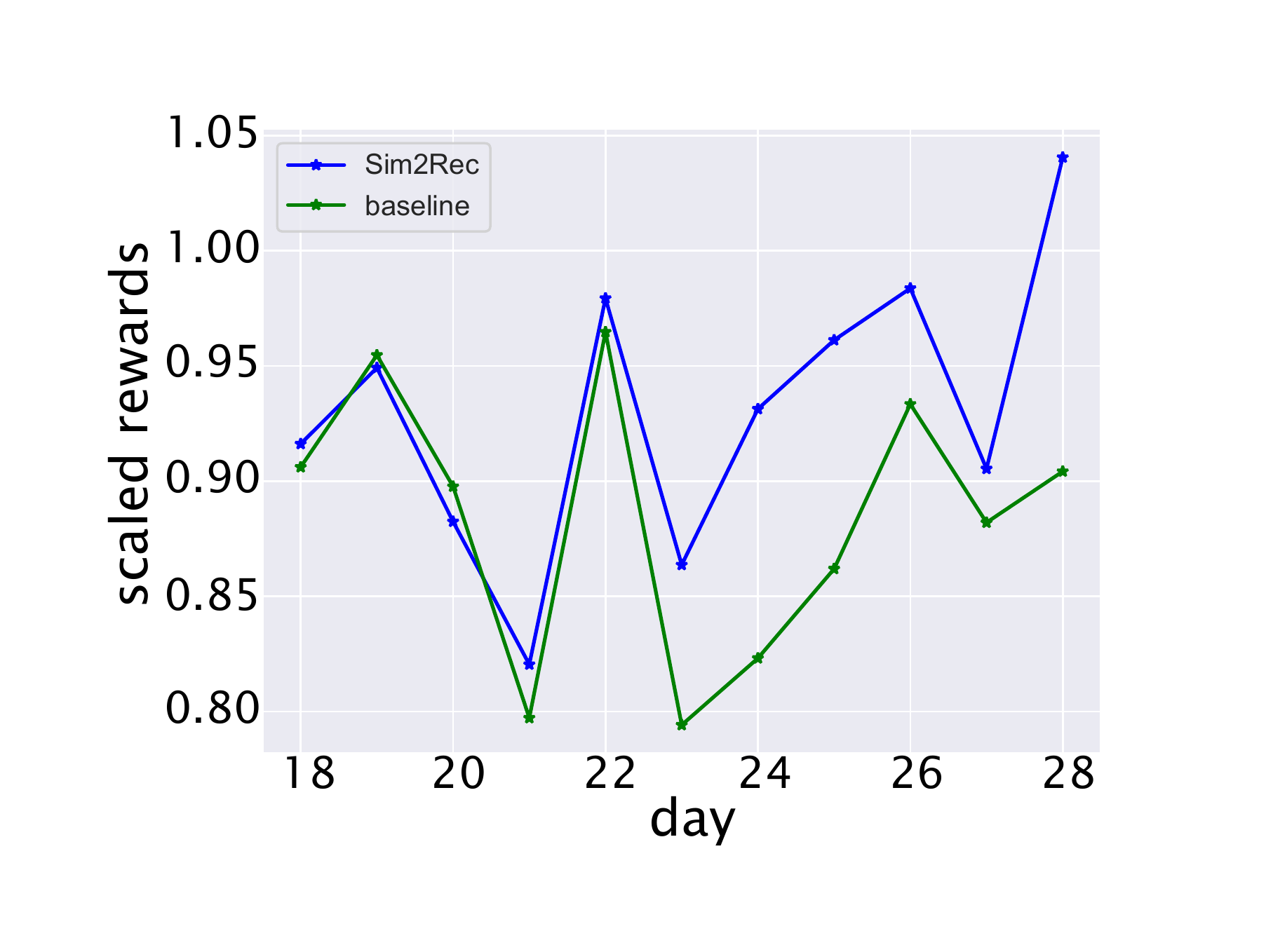}
	    \vspace{-2mm}
    	\caption{Illustration of the online test. The X-axis is the date. The Y-axis is the average daily reward. }
    	\label{fig:demer-daily-fos}
	    \vspace{-2mm}
    \end{figure}

	We finally deploy the policy trained by \our{} to the real world and test the performance for 7 days. The baseline is a simulator-based method, DR-UNI, which implement with the same simulator set and RL algorithm~\citep{ppo} as Sim2Rec  without the extractor and context-aware policy. 
	The results are shown in~Fig.~\ref{fig:demer-daily-fos}. We split the drivers into control and treatment groups and deploy the policy from day 22 to day 28 of a month in the treatment group. Before deployed, drivers are recommended with the same human policy. We find that the performance improvement of the baseline policy is 0.1\%, which is similar to the performance before the AB Test, while the improvement of \our{} is 6.9\%, which is significantly better than the human policy and baseline policy.

